\documentclass[]{aastex631}

\newcommand{\LiCaSSP}{$\log(\text{Li/Ca})_{\text{SSP}}$}
\newcommand{\NaCaSSP}{$\log(\text{Na/Ca})_{\text{SSP}}$}
\newcommand{\MgCaSSP}{$\log(\text{Mg/Ca})_{\text{SSP}}$}
\newcommand{\KCaSSP}{$\log(\text{K/Ca})_{\text{SSP}}$}
\newcommand{\FeCaSSP}{$\log(\text{Fe/Ca})_{\text{SSP}}$}
\newcommand{\CrCaSSP}{$\log(\text{Cr/Ca})_{\text{SSP}}$}

\newcommand{\Teff}{$T_{\text{eff}}$}

\shorttitle{Origins of WD Lithium Enhancement}
\shortauthors{Kaiser et al.}
\graphicspath{{./}{figures/}}
\begin{document}

\title{The Origins of Lithium Enhancement in Polluted White Dwarfs}

\correspondingauthor{Benjamin C. Kaiser}
\email{ben.kaiser@unc.edu}

\author[0000-0003-1970-4684]{Benjamin C. Kaiser}
\affiliation{Department of Physics and Astronomy, University of North Carolina, 
120 E. Cameron Ave., 
Chapel Hill, NC 27599, USA}

\author{J. Christopher Clemens}
\affiliation{Department of Physics and Astronomy, University of North Carolina,
120 E. Cameron Ave.,
Chapel Hill, NC 27599, USA}

\author[0000-0002-9632-1436]{Simon Blouin}
\affiliation{Department of Physics and Astronomy, University of Victoria,Victoria
BC V8W 2Y2, Canada}

\author[0000-0003-2852-268X]{Erik Dennihy}
\affiliation{Gemini Observatory/NSF's NOIRLab, Casilla 603, La Serena, Chile}
\affiliation{Rubin Observatory Project Office, 950 N. Cherry Avenue, Tucson, AZ 85719, USA}

\author[0000-0003-4609-4500]{Patrick Dufour}
\affiliation{D\'epartement de Physique, Universit\'e de Montr\'eal, Montr\'eal, QC H3C 3J7, Canada}
\affiliation{Institut de Recherche sur les Exoplan\'etes (iREx), Universit\'e de Montr\'eal, Montr\'eal, QC H3C 3J7, Canada}

\author[0000-0003-4145-3770]{Ryan J. Hegedus}
\affiliation{Department of Physics and Astronomy, University of North Carolina,
120 E. Cameron Ave.,
Chapel Hill, NC 27599, USA}

\author[0000-0003-1862-2951]{Joshua S. Reding}
\affiliation{Department of Physics and Astronomy, University of North Carolina,
120 E. Cameron Ave.,
Chapel Hill, NC 27599, USA}

%\altaffiliation{AASTeX v6+ programmer}

%\affiliation{AAS Director of Publishing}
%\affiliation{American Astronomical Society \\
%1667 K Street NW, Suite 800 \\



\begin{abstract}
The bulk abundances of exoplanetesimals can be measured when they are accreted by white dwarfs. Recently, lithium from the accretion of exoplanetesimals was detected in relatively high levels in multiple white dwarfs.
There are presently three proposed hypotheses to explain the detection of excess lithium in white dwarf photospheres: Big Bang and Galactic nucleosynthesis, continental crust, and an exomoon formed from spalled ring material.
We present new observations of three previously known lithium-polluted white dwarfs (WD~J1824+1213, WD~J2317+1830, and LHS~2534), and one with metal pollution without lithium (SDSS~J1636+1619). We also present atmospheric model fits to these white dwarfs.
%and two other previously known, lithium-polluted white dwarfs (WD~J1644–0449 and SDSS~J1330+6435).
We then evaluate the abundances of these white dwarfs and two additional lithium-polluted white dwarfs that were previously fit using the same atmospheric models (WD~J1644–0449 and SDSS~J1330+6435) in the context of the three extant hypotheses for explaining lithium excesses in polluted white dwarfs. We find Big Bang and Galactic nucleosynthesis to be the most plausible explanation of the abundances in WD~J1644–0449, WD~J1824+1213, and WD~J2317+1830. SDSS~J1330+6435 will require stricter abundances to determine its planetesimal's origins, and LHS~2534, as presently modeled, defies all three hypotheses. We find the accretion of an exomoon formed from spalled ring material to be highly unlikely to be the explanation of the lithium excess in any of these cases.

\end{abstract}


\section{Introduction}\label{sec:intro}

White dwarfs are the end state of stellar evolution for the vast majority of stars. White dwarfs' high surface gravity causes elements to stratify by mass, so in isolation, only the lightest elements, H and He, should be detectable at the surface.\footnote{C can be detected as a result of dredge up at lower temperatures in some white dwarfs \citep[e.g.][]{Fontaine1984_Cdredgeup} and other metals can be detected in hotter white dwarfs due to radiative levitation \citep[e.g.][]{Bruhweiler1983_radiativelevitation}.} However, hundreds of white dwarfs have had heavier elements detected in their photospheres. These heavier elements are from recently accreted extrasolar planetary material, and their presence in the white dwarf photosphere enables a measurement of their relative abundances. Thus, we can obtain bulk abundance measurements of extrasolar planetesimals.

Dozens of different elements have been detected from accreted extrasolar planetesimals in white dwarfs at various relative abundances (see \citet{Klein2021_BeWD} for a review). The differing abundance ratios have largely been attributed to the diversity of planetary abundances with the majority being comparable to Solar System bodies or parts thereof \citep{Hollands2018_DZII,Bonsor2020_differentiation}. In 2021, lithium was discovered in five polluted white dwarfs \citep{Kaiser2021_LiWD, Hollands2021_LiWD}: WD~J164417.01$-$044947.7 (hereafter WD~J1644–0449)\footnote{We provide the full J2000 white dwarf name first, but we refer to each white dwarf throughout this work using a shorter name that was originally used in the literature.} \citep{Kaiser2021_LiWD}, WD~J133001.17+643523.69 (hereafter SDSS~J1330+6435) \citep{Kaiser2021_LiWD,Hollands2021_LiWD}, WD~J121456.38$-$023402.84 (hereafter LHS~2534) \citep{Hollands2021_LiWD}, WD~J182458.45+121316.82 (hereafter WD~J1824+1213) \citep{Hollands2021_LiWD}, and WD~J231726.74+183052.75 (hereafter WD~J2317+1830) \citep{Hollands2021_LiWD}. All five white dwarfs' lithium abundance ratios—Li/Ca in \citet{Kaiser2021_LiWD} and Li/Na in \citet{Hollands2021_LiWD}—were determined to be elevated compared to CI Chondrites from the Solar System. \citet{Kaiser2021_LiWD} proposed the lithium enhancement is a consequence of nucleosynthetic evolution of the Galaxy in which Big Bang nucleosynthesis (BBN) provides higher Li/Ca in the early gas. \citet{Hollands2021_LiWD} proposed the lithium enhancement is a result of the accretion of continental crust with one of the white dwarfs accreting a ``particularly lithium-rich'' piece of continental crust. 

Another light element, Be, was recently discovered in two white dwarf photospheres \citep{Klein2021_BeWD}, and its abundance relative to Ca was also elevated relative to CI Chondrites. \citet{Klein2021_BeWD} posited the Be-enhancement was a consequence of the accreted planetesimals' presence in a high proton bombardment location. \citet{Doyle2021_spallWD} proposed the high proton bombardment location as the icy rings of a gas giant with a magnetic field and proton flux comparable to Jupiter's. The accreted bodies would be icy exomoons formed from former ring material that had been subjected to the high proton flux, which caused the spalling of O to produce elevated levels of Be. \citet{Doyle2021_spallWD} pointed out that elevated Li levels would also result from this spallation, and suggested this could be relevant for the Li-enhanced white dwarfs as well. Thus, there are three extant hypotheses to explain elevated Li levels of planetesimals accreted by white dwarfs:
\begin{enumerate}
\item Galactic nucleosynthetic evolution caused a dearth of heavier metals relative to lithium at early Galactic epochs \citep{Prantzos2012_Gal_LiBeB}. Correspondingly, metal-deficient $-$and therefore Li-enhanced$-$ exoplanetesimals formed from these nebulae with the white dwarf progenitors. These Li-enhanced planetesimals were then accreted by the white dwarfs causing the observed pollution \citep{Kaiser2021_LiWD}.
\item The continental crust of the Earth is enhanced in Li and K compared to CI Chondrites \citep{Rumble2019_contcrust}. The white dwarfs accreted ``particularly lithium-rich'' continental crust material, which was presumably formed on Earth-like exoplanets that orbited the white dwarf progenitors \citep{Hollands2021_LiWD}. 
\item Li, Be, and B can be produced by bombarding C, N, and O nuclei with high energy protons in a process called spallation \citep{Read1984_nukecross}. Icy exoplanetary rings around Jupiter-like gas giants were spalled by protons captured from the white dwarf progenitors' stellar wind by the gas giants' magnetospheres. This nuclear spallation produced Li, Be, and B in the exoplanetary rings. The rings were then reaccumulated to form exomoons, which were accreted by the white dwarfs \citep{Doyle2021_spallWD}.
\end{enumerate}

We evaluate the abundances of the bodies accreted by the 5 Li-polluted whited dwarfs in the context of these 3 hypotheses, with the same atmospheric models used to derive all abundances consistently. First, we present follow-up spectroscopic observations aimed at assessing these three hypotheses in Section \ref{sec:obs}. Our new observations include the following white dwarfs with Li identifications: LHS~2534, WD~J1824+1213, and WD~J2317+1830. We also present Gemini GMOS-N spectroscopic observations of the low temperature DZ WD~J163601.35+161907.51 (hereafter SDSS~J1636+1619) with null detections of Li and K. We perform atmospheric model fits to the Li-polluted white dwarfs and SDSS~J1636+1619 in Section \ref{sec:atm_modeling}. We infer the abundances of the extrasolar planetesimals accreted by the 5 white dwarfs' with Li detections and SDSS~J1636+1619 in Section \ref{sec:infer_abund}. We compute total ages of the white dwarfs discussed in Section \ref{sec:WD_ages}. We evaluate the kinematic population memberships of the white dwarfs in Section \ref{sec:kinematics}. We derive the expected Galactic nucleosynthetic evolution of various abundances from archival data in Section \ref{sec:saga_abunds}. We address the spallation hypothesis in Section \ref{sec:spallation}. We evaluate each white dwarf in the context of the geologic differentiation hypothesis and Big Bang combined with Galactic nucleosynthesis hypothesis in Section \ref{sec:nuke_vs_geo}. In Section \ref{sec:caveats}, we discuss additional caveats, and we summarize our findings in Section \ref{sec:conclusions}. 

\section{Observations and Reductions}\label{sec:obs}
\begin{table}[!ht]
    \centering
    \begin{tabular}{|l|l|l|l|l|l|l|}
    \hline
        Object & Obs. Date & Telescope & Instrument/Setup & Res. Power & Source & Use in Modeling  \\ \hline
        LHS~2534  & 2019-01-14 & VLT & X-Shooter & 8,200 & \citet{Hollands2021_LiWD} & Li, Na, K, Ca, Cr, Fe  \\ 
        ~ & 2021-01-09 & SOAR & Goodman/400M1 & 700 & This work & ---  \\ 
        ~ & 2021-01-09 & SOAR & Goodman/400M2 & 700 & This work & ---  \\ \hline
%        SDSS~J1330+6435 & ~ & SDSS & SDSS & ~ & SDSS DR7 & Ca, Na, Li, K-limit  \\ \hline
        %SDSS~J1636+1619 & ~ & SDSS & ~ & ~ & ~ & Ca  \\
        SDSS~J1636+1619 & 2020-07-24 & Gemini-N & GMOS-N & 1,000 & This work & Li-limit, Na-limit, K-limit  \\ \hline
        %WD~J1644$-$0449  & 2019-07-03 & SOAR & Goodman & 400M2 & \citet{Kaiser2021_LiWD} & ---  \\ 
        %~ & 2019-07-29 & SOAR & Goodman & 400M1 & \citet{Kaiser2021_LiWD} & Ca, Na, Li, MgH  \\
        %~ &  2019-08-25 & SOAR & Goodman & 400M2 & \citet{Kaiser2021_LiWD} & K  \\ \hline
        WD~J1824+1213  & 2018-08-7/8 & WHT & ISIS & 3,500 & \citet{Hollands2021_LiWD} & Li, Na, Ca  \\ 
        ~ & 2021-06-11 & SOAR & Goodman/400M1 & 700 & This work & MgH  \\
        ~ &  2021-06-11 & SOAR & Goodman/400M2 & 700 & This work & K-limit  \\ \hline
        WD~J2317+1830  & 2018-09-02 & GTC & OSIRIS & 900 & \citet{Hollands2021_LiWD} & Li, Na,  K-limit, Ca \\
        ~ &  2021-07-04 & SOAR & Goodman/400M2 & 700 & This work & --- \\ \hline
    \end{tabular}
    
    \caption{Observations presented in this work or used in the atmospheric modeling presented in this work. Some of these spectra were modeled by \citet{Hollands2021_LiWD}, but we model them in this work with the models of \citet{Blouin2018_PaperI} to create a homogeneous set of abundances.\label{tab:obs}}
\end{table}

\subsection{SOAR/Goodman Spectroscopy}\label{sec:soar}

\begin{figure*}[htb!]
\includegraphics[scale=1]{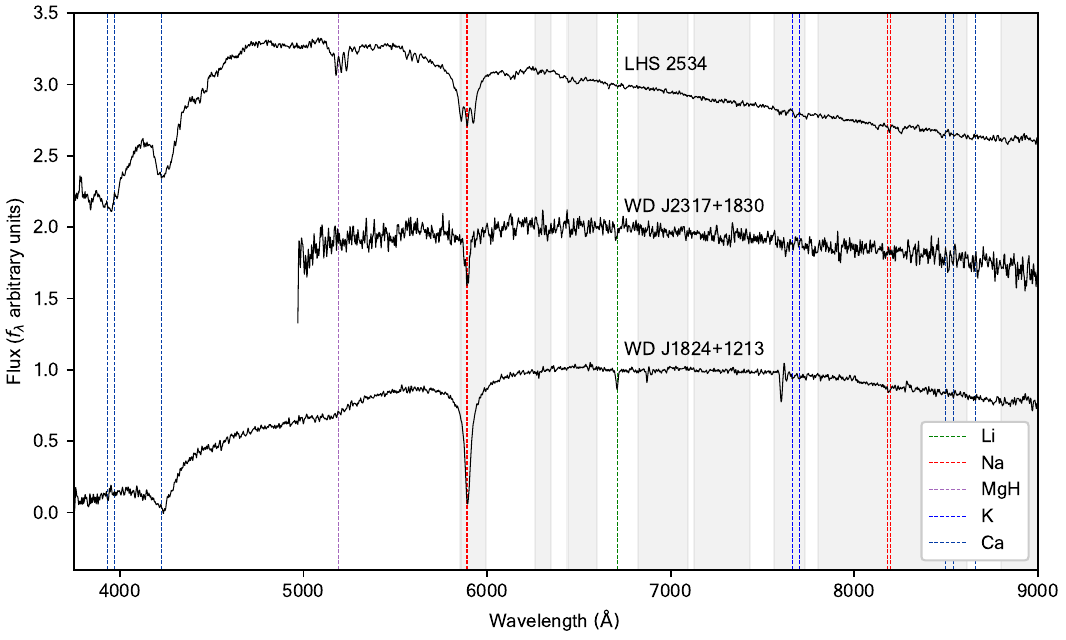}

\centering
\caption{SOAR/Goodman High Throughput Spectrograph Spectra smoothed with a 3-pixel boxcar. The 400M1 and 400M2 spectra of LHS~2534 and WD~J1824+1213 were stitched together at 6660 $\text{\AA}$. Line markers are plotted as well, but not all spectra exhibit all lines. Regions of telluric absorption are shaded in grey. WD~J1824+1213 displays a telluric removal artifact to the blue side of the \ion{K}{1} resonance line wavelengths. \label{fig:goodman_spectra}}
\end{figure*}

The new spectra spectra we present in this work for WD~J1824+1213, WD~J2317+1830, and LHS~2534 were obtained using the Goodman Spectrograph \citep{Clemens2004_Goodman} on the Southern Astrophysical Research (SOAR) Telescope. The targets were observed using the 400 l/mm grating and 3.2" slit in the M1 and/or M2 spectroscopic setups (hereafter 400M1 and 400M2), which yielded a seeing-dependent resolving power $\lambda/\Delta \lambda \approx 700$. The 400M1 spectra cover 3740–7095~Å, and the 400M2 spectra cover 4995–9020~Å. Full details of the individual observations are in Table \ref{tab:obs}. Reductions were performed using a custom Python reduction pipeline \citep{Kaiser2021_LiWD}. The \ion{K}{1} resonance lines are located on the edge of a telluric $\text{O}_2$ band, so we performed telluric corrections on the 400M2 spectra in order to probe for the presence of K. The corrections were performed by dividing the target spectrum by the telluric absorption of a spectrophotometric standard observed at similar airmass on the same observing night. The 400M2 spectra of LHS~2534 and WD~J1824+1214 were corrected for telluric absorption using observations of the spectrophotometric standard LTT 3218 while the 400M2 spectrum of WD~J2317+1830 was corrected using the spectrophotometric standard LTT 7987 using the reference spectra from \citet{Moehler2014_fluxcal}. Our 400M1 spectrum of WD~J1824+1213 filled a wavelength gap in the spectrum from \citet{Hollands2021_LiWD} described in Section \ref{sec:archival}, which allowed for the detection and modeling of MgH. The telluric-corrected 400M2 spectrum of WD~J1824+1213 provided the new K limit. The spectra are presented in Figure \ref{fig:goodman_spectra}.

\subsection{Gemini/GMOS-N Spectroscopy}\label{sec:gmos}

\begin{figure*}[htb!]
\includegraphics[scale=1]{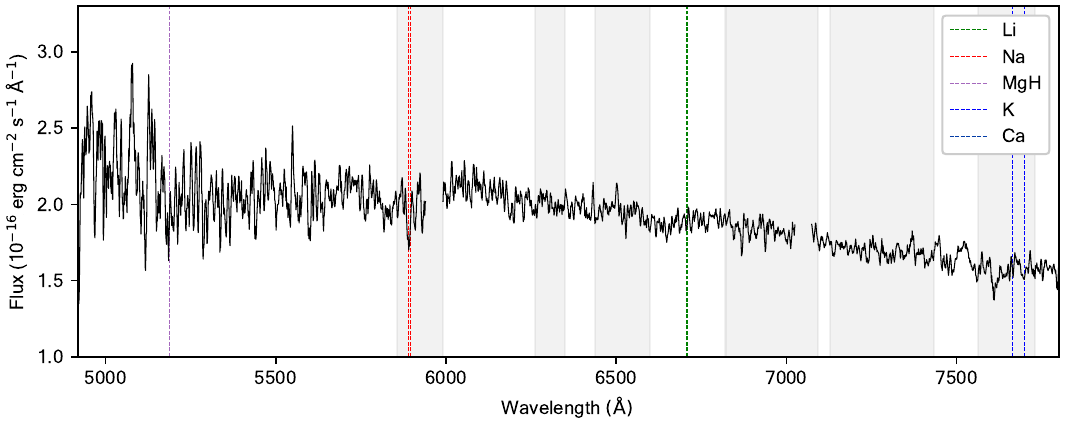}

\centering
\caption{Gemini-North/GMOS-N spectrum of SDSS~J1636+1619 smoothed with a 5-pixel boxcar. There are detector gaps present in the spectrum near 5900 $\text{\AA}$ and 7100 $\text{\AA}$. No absorption lines are detected, but line markers are placed where lines would be. \label{fig:GMOS_J1636}}
\end{figure*}

We targeted SDSS~J1636+1619 because it was one of the few extremely low-$T_{\text{eff}}$ DZs in the Montreal White Dwarf Database \citep{Dufour2017_MWDD} at the time of observation, and all of the Li detections were in low-$T_{\text{eff}}$ DZs. We observed SDSS~J1636+1619 using the Gemini Multi-Object Spectrograph-North (GMOS-N) \citep{Hook2004_GMOS-N} mounted on the Gemini-North telescope in a Director's Discretionary Time allocation (GN-2020A-DD-114, PI: Kaiser) on the night of 2020 July 24. We used the 1.0" slit with the B600 grating centered at 6500 $\text{\AA}$ and the GG455 order blocking filter, resulting in a wavelength coverage of 4914–8131 $\text{\AA}$ and a central wavelength resolving power of $\lambda/\Delta\lambda \approx 1,000$. The position angle was set to the parallactic angle to minimize wavelength-dependent slit losses.
The GMOS data were processed using a combination of the Gemini IRAF package and an optimal extraction routine based on the methods described in \citet{Marsh1989_specextraction}.
A telluric standard star (EG 131) was observed immediately afterwards at the same airmass in the same setup. This standard was used for flux calibration and telluric corrections in the same way as described in Section \ref{sec:soar}.
The reference spectrum was from \citet{Bessell1999_EG131}.
As in \citet{Kaiser2021_LiWD}, we also used the extinction curve of \citet{Stritzinger2005} for Cerro Tololo despite these observations occurring at Mauna Kea.
We therefore are not confident in our overall flux calibration, but we are confident in our telluric corrections. Additionally, the overall flux calibration is unimportant for determining abundance limits in this context. The reduced spectrum is presented in Figure \ref{fig:GMOS_J1636}. There are no metals lines present in the GMOS-N spectrum; Li was not detected.

\subsection{Archival Spectroscopy}\label{sec:archival}
As none of the white dwarfs analyzed in this work are new discoveries, there exist archival spectroscopic data. In the interest of performing the most robust analysis of abundances, we incorporate data from past works. In the cases of WD~J1824+1213, WD~J2317+1830, and LHS~2534, \citet{Hollands2021_LiWD} presented spectra with superior resolution and signal-to-noise compared to the newly collected Goodman spectroscopy, so we used those whenever possible in our analysis. \citet{Hollands2021_LiWD} obtained the LHS~2534 spectrum using X-Shooter on the Very Large Telescope (VLT) and corrected the telluric absorption using MOLECFIT \citep{Smette2015_Molecfit_I,Kausch2015_Molecfit_II}. The WD~J2317+1830 spectrum of \citet{Hollands2021_LiWD} was obtained using the Optical System for Imaging and low-Intermediate-Resolution Integrated Spectroscopy (OSIRIS) on the Gran Telescopio Canarias (GTC) and originated from the 40 pc survey of \citet{Tremblay2020_40pcnorthspectro}. \citet{Hollands2021_LiWD} attempted telluric removal on the WD~J2317+1830 spectrum using an observation of a standard star, but it was observed using a wider slit, so the telluric removal was not as successful. The WD~J1824+1213 spectrum of \citet{Hollands2021_LiWD} was obtained using the Intermediate-dispersion Spectrograph and Imaging System (ISIS) on the William-Herschel Telescope (WHT) and also originated from the 40 pc survey of \citet{Tremblay2020_40pcnorthspectro}. We obtained the fully reduced versions of these spectra from \citet{Hollands2021_LiWD}, and we used them for atmospheric modeling (see Section \ref{sec:atm_modeling}).

\section{Analysis}\label{sec:analysis}

\subsection{Atmospheric Modeling}\label{sec:atm_modeling}
The white dwarf atmospheric parameters and abundances used in this work were all derived using the model atmospheres of \citet{Blouin2018_PaperI,Blouin2018_paperII,Blouin2019_paperIII}, specifically tailored for low-$T_{\rm eff}$ white dwarfs. These models include a detailed treatment of high-density effects on the spectral lines, continuum opacities, and equations of state.
Our fitting procedure employs an iterative approach that combines both photometric and spectroscopic data to achieve self-consistent solutions \citep[e.g.,][]{Coutu2019_lowteffwds}. The process begins by adjusting the model’s effective temperature, solid angle $\pi R^2/D^2$, and hydrogen abundance to match the available photometry. Using {\it Gaia} eDR3 parallaxes \citep{Gaia2021_GaiaeDR3astrometry}, we derive the white dwarf radius, which in turn allows for the calculation of mass and surface gravity using the evolutionary models of \citet{Bedard2022_specevoII}. With these initial $T_{\rm eff}$, $\log g$ and $\log\,{\rm H/He}$ values, we then fit the spectroscopic data to determine metal abundances. After adjusting the metal abundances, we return to the photometric fit, now incorporating the new abundance values. This process of alternating between photometric and spectroscopic fits is repeated until the atmospheric parameters converge to a stable solution.

For WD~J1824+1213, we used Pan-STARRS photometry \citep{Chambers2016_panstarrs} supplemented by UKIDSS J \citep{Lawrence2007_UKIDSSsurvey} and WISE W1, W2 bands \citep{Cutri2014_allwisecatalogue}. We modeled WD~J2317+1830 using Pan-STARRS photometry \citep{Chambers2016_panstarrs} and UKIDSS J-band data \citep{Lawrence2007_UKIDSSsurvey}. For LHS 2534, we employed SDSS photometry \citep{SDSS2020_DR16specandphoto} along with 2MASS J, H \citep{Skrutskie2006_2MASSsurvey} and WISE W1, W2 bands \citep{Cutri2014_allwisecatalogue}. These photometric datasets were combined with spectra from \citet{Hollands2021_LiWD} as the primary spectroscopic data, supplemented by our newly obtained Goodman spectra to fill wavelength gaps. As a consistency check, we also performed independent fits using only the Goodman spectra. The results for overlapping wavelength regions were found to be consistent across instruments.

Some spectral lines of WD~J1824+1213 and LHS 2534 are too narrow compared to the models' predictions, an issue already noted by \citet{Hollands2021_LiWD}. The origin of this problem remains unknown, and for the affected lines we resorted to comparing equivalent widths from the observations to equivalent widths from the models to measure the metal abundances. 
SDSS~J1636+1619 was modeled using previous fits to SDSS photometry and spectroscopy from \citet{Blouin2020_MgcoolWD}, with the addition of our new Gemini GMOS-N spectrum to derive abundance limits for Li, Na, and K. The other atmospheric parameters for this object were retained from \citet{Blouin2020_MgcoolWD}, as they were derived using the same underlying models as the other objects.

WD~J1644–0449 and SDSS~J1330+6435 parameters were taken from \citet{Kaiser2021_LiWD}. However, for WD~J1644–0449, we recalculated the mass using the updated parallax from {\it Gaia} eDR3 \citep{Gaia2021_GaiaeDR3astrometry}, which showed an increase of $\sim 5$\% compared to the {\it Gaia} DR2 value. This revision resulted in an increase in the estimated mass from $0.45 \pm 0.12 M_{\odot}$ as reported in \citet{Kaiser2021_LiWD} to $0.49 \pm 0.13 M_{\odot}$ in the current analysis. While these two mass estimates are consistent within their uncertainties, the updated central value of the mass distribution affects the total age calculation (see Section \ref{sec:WD_ages}).

Table \ref{tab:atm_params} presents the best-fit atmospheric parameters for all six white dwarfs, indicating which parameters were derived in this work and which were taken from previous studies using the same modeling framework. Figures of the new model fits to spectroscopy and photometry are included in Appendix \ref{sec:model_fit_plots} with the exception of SDSS~J1636+1619 because the new spectroscopy did not exhibit any absorption lines.

%\begin{splitdeluxetable*}{lLRRRBRRRRRRR}
\begin{splitdeluxetable*}{lLCCCBCCCCCCC}
\tablecaption{Best-fit atmospheric parameters of the white dwarfs considered in this work. All parameters are derived using the same models, but not all parameters are derived in this work. See Section \ref{sec:atm_modeling}.\label{tab:atm_params}}
\tablewidth{0pt}
\tablehead{
\colhead{Object} & \colhead{$T_{\text{eff}}$ (K)} & \colhead{log(g)} & \colhead{$\text{M}_{\text{WD}}$ ($\text{M}_{\odot}$)} & \colhead{log(H/He)} & \colhead{log(Li/He)} & \colhead{log(Na/He)} & \colhead{log(Mg/He)} & \colhead{log(K/He)} & \colhead{log(Ca/He)} & \colhead{log(Cr/He)} & \colhead{log(Fe/He)}
}
\startdata
WD~J1644$-$0449 & 3830\pm230\tablenotemark{d} & 7.85\pm0.23 & 0.49\pm0.13 & <-2.0\tablenotemark{d} & -11.2\pm0.2\tablenotemark{d} & -9.5\pm0.2\tablenotemark{d} & ~& -10.9\pm0.2\tablenotemark{d} & -9.5\pm0.2\tablenotemark{d} & ~ &  ~ \\ 
SDSS~J1330+6435 & 4310\pm190\tablenotemark{b} & 8.26\pm0.15\tablenotemark{b} & 0.74\pm0.10\tablenotemark{b} & ~ & -10.3\pm0.2\tablenotemark{d} & -8.5\pm0.3\tablenotemark{b} & ~ & <-9.1\tablenotemark{d} & -8.8\pm0.3\tablenotemark{b} & ~ & ~  \\ 
WD~J1824+1213 & 3540\pm90 & 7.53\pm0.09 & 0.33\pm0.04 & -0.31\pm0.25 & -11.84\pm0.10 & -10.18\pm0.12 & -8.98\pm0.30 & <-11.4 & -9.80\pm0.20 & ~  &~   \\ 
WD~J2317+1830 & 4430\pm120 & 8.74\pm0.06 & 1.05\pm0.03 & -0.2\pm0.2 & -10.60\pm0.12 & -9.65\pm0.10 & ~ & <-9.5 & -10.42\pm0.14 & ~ & ~  \\ 
LHS~2534 & 5020\pm100 & 8.10\pm0.08 & 0.64\pm0.05 & <-3.0 & -11.03\pm0.12 & -8.96\pm0.08 & ~ & -9.41\pm0.12 & -9.62\pm0.10 & -9.76\pm0.10 & -8.58\pm0.20  \\ 
%WD~J2356$-$209 & 4040\pm110\tablenotemark{b} & 7.98\pm0.07\tablenotemark{b} & 0.56\pm0.04\tablenotemark{b} & -1.5\pm0.2\tablenotemark{b} & <-11.7\tablenotemark{d} & -8.3\pm0.2\tablenotemark{b} & -8.0\pm0.2\tablenotemark{b} & <-10.4\tablenotemark{d} & -9.4\pm0.1\tablenotemark{a} & <-9.9\tablenotemark{b} & -8.6\pm0.2\tablenotemark{b}  \\
SDSS~J1636+1619 & 4410\pm200\tablenotemark{c} & 8.10\pm0.06\tablenotemark{c} & 0.67\pm0.04\tablenotemark{c} & <-3.0\tablenotemark{c} & <-11.3 & <-9.9 & <-8.3\tablenotemark{c} & <-9.9 & -9.5\pm0.1\tablenotemark{c} & ~ & <-8.3\tablenotemark{c} \\
\enddata
\tablenotetext{a}{\citet{Blouin2019_CaI}}
\tablenotetext{b}{\citet{Blouin2019_paperIII}}
\tablenotetext{c}{\citet{Blouin2020_MgcoolWD}}
\tablenotetext{d}{\citet{Kaiser2021_LiWD}}

\end{splitdeluxetable*}

\subsection{Inferring Abundances of Accreted Material}\label{sec:infer_abund}

White dwarf accretion of extrasolar planetesimals is somewhat simplistically treated as occurring in three phases: increasing, steady-state, and decreasing \citep{Koester2009_difftimes,Harrison2018_geoexomaterial,Swan2019_exoplanets}. The abundances of the accreted body will appear differently in the photosphere depending upon the accretion phase, which means the abundances we infer for the accreted body depend upon the assumed accretion phase.

\subsubsection{Increasing Phase}\label{sec:increasing}
\begin{figure*}[htb!]
\includegraphics[scale=1]{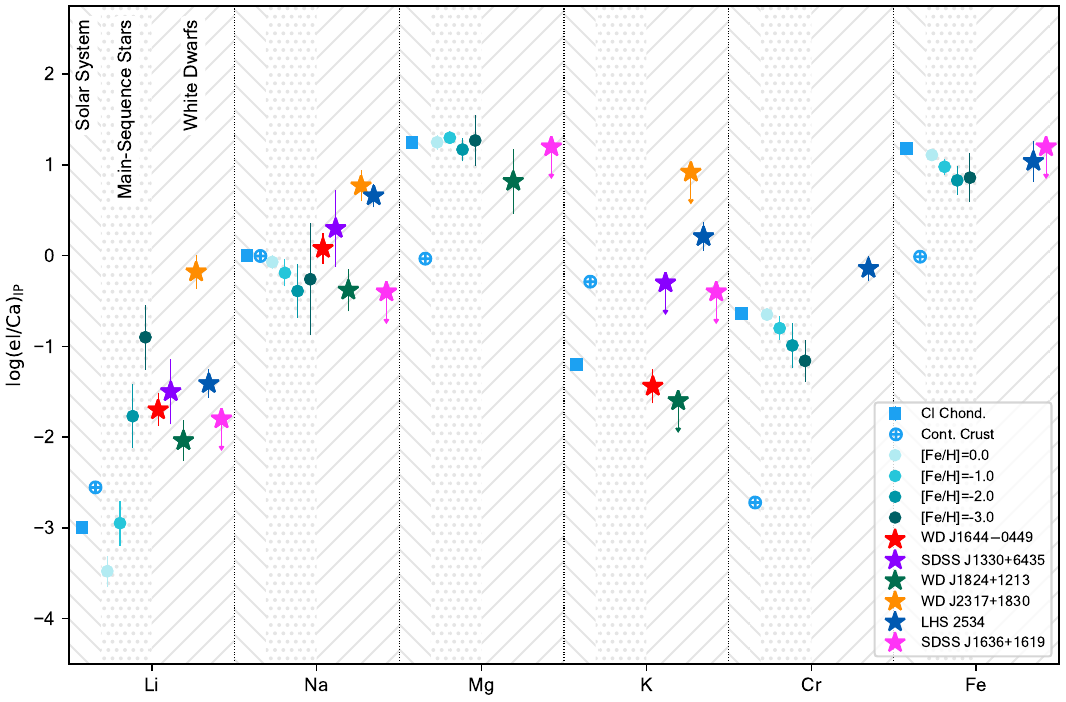}

\centering
\caption{Exoplanetary pollutant material abundances relative to Ca as measured in each white dwarf assuming increasing phase pollution (stars). The inferred increasing phase abundances are the same as the photospheric abundances (see Section \ref{sec:increasing}). The abundances of CI Chondrites (blue square) \citep{Lodders2019_solarabund} and Earth's Continental Crust (blue cross-dot) \citep{Rumble2019_contcrust} are plotted for comparison. The abundances of main-sequence stars from the SAGA database \citep{Suda2017_SAGAIV} in 1-dex wide [Fe/H] bins centered on 0, –1, –2, and –3 are plotted as solid circles. We determined the log(Li/Ca) abundances for the main-sequence stars using Equation \ref{eq:infer_saga_LiCa} rather than taking the SAGA database measured log(Li/Ca) abundances because Li abundances from main-sequence stars are underabundant compared to their protostellar nebulae \citep{Fu2015_Limetalpoorstars}. Different hatched regions are included to aid the eye. White dwarf IP abundances are provided in Table \ref{tab:IP_abunds}. \label{fig:allel_vs_Ca_phot}}
\end{figure*}

In the increasing phase (IP), accretion is causing the amount of planetary material in the white dwarf photosphere to increase.
%The relative abundances of the various heavy elements in the photosphere (e.g. Li/Ca) directly reflect the relative abundances in the accreted planetary material.
Diffusion is occurring at a much slower rate than accretion, so the relative abundances of the elements (e.g. Li/Ca) are the same in the photosphere as they are in the accreted planetary material. If the accretion is in the increasing phase, the parent planetary body abundances are obtained by taking the photospheric abundances:

\begin{equation}\label{eq:inc_phase}
\log(\text{el}_1/\text{el}_2)_{\text{IP}}=\log(\text{el}_1/\text{el}_2)_{\text{Phot.}},
\end{equation}

where $\log(\text{el}_1/\text{el}_2)_{\text{IP}}$ is the inferred abundance of the accreted body assuming increasing phase, and $\log(\text{el}_1/\text{el}_2)_{\text{Phot.}}$ is the measured abundance in the white dwarf photosphere as derived in Section \ref{sec:atm_modeling} \citep{Harrison2018_geoexomaterial}. The abundances and error bars for the accreted body assuming increasing phase are given in Table \ref{tab:IP_abunds}. The abundances are plotted as stars in Figure \ref{fig:allel_vs_Ca_phot}.

\begin{deluxetable*}{lRRRRRR}[!ht]
\tablecaption{Inferred abundances of the accreted planetesimals assuming accretion is in the increasing phase. These are equal to the photospheric abundances as shown in Equation \ref{eq:inc_phase}.\label{tab:IP_abunds}}
\tablewidth{0pt}
\tablehead{\colhead{Object} & \colhead{$\log(\text{Li}/\text{Ca})_{\text{IP}}$} & \colhead{$\log(\text{Na}/\text{Ca})_{\text{IP}}$} & \colhead{$\log(\text{Mg}/\text{Ca})_{\text{IP}}$} & \colhead{$\log(\text{K}/\text{Ca})_{\text{IP}}$} & \colhead{$\log(\text{Cr}/\text{Ca})_{\text{IP}}$} & \colhead{$\log(\text{Fe}/\text{Ca})_{\text{IP}}$}
}
    \startdata
        WD~J1644$-$0449 & -1.70\pm0.18 & 0.08\pm0.17 & ~ & -1.44\pm0.19 & ~ & ~  \\ 
        SDSS~J1330+6435 & -1.50\pm0.36 & 0.30\pm0.42 & ~ & <-0.3 & ~ & ~  \\ 
        WD~J1824+1213 & -2.04\pm0.22 & -0.38\pm0.23 & 0.82\pm0.36 & <-1.6 & ~ & ~  \\
        WD~J2317+1830 & -0.18\pm0.18 & 0.77\pm0.17 & ~ & <0.9 & ~ & ~  \\ 
        LHS~2534 & -1.41\pm0.16 & 0.66\pm0.13 & ~ & 0.21\pm0.16& -0.14\pm0.14 & 1.04\pm0.22 \\ 
        %WD~J2356$-$209 & <-2.3 & 1.1\pm0.2 & 1.4\pm0.22& <-0.95 & <-0.5 & 0.8\pm0.22 \\
        SDSS~J1636+1619 & <-1.8 & <-0.4 & <1.2 & <-0.4 & ~ & <1.2\\ 
    \enddata
\end{deluxetable*}

\subsubsection{Steady-State Phase}\label{sec:steadystate}

\begin{figure*}[htb!]
\includegraphics[scale=1]{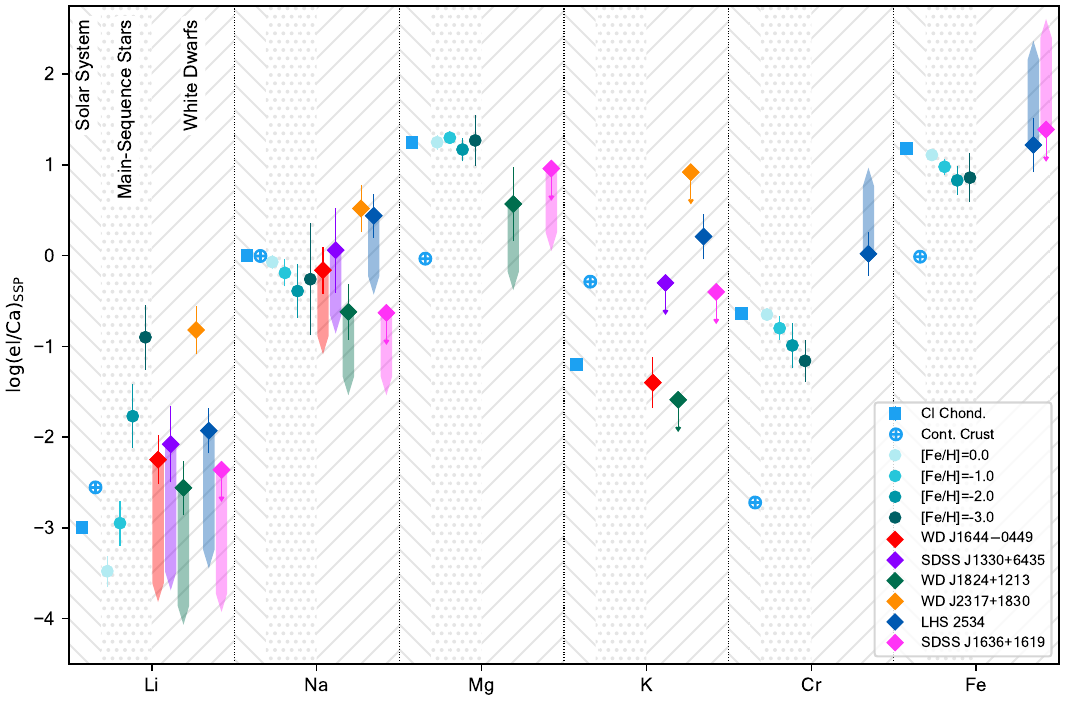}

\centering
\caption{Exoplanetary pollutant material abundances relative to Ca as measured in each white dwarf assuming steady-state phase pollution are represented by diamonds. The inferred steady-state phase abundances are calculated using Equation \ref{eq:ss_phase} (see Section \ref{sec:steadystate}). The long, faded arrows point to the abundances of the accreted bodies if accretion is 5~$\tau_{\text{Ca}}$ into the decreasing phase following steady-state accretion. WD~J2317+1830 (orange diamond) lacks decreasing phase arrows because its accretion is believed to be ongoing as originally inferred by \citet{Hollands2021_LiWD}. The decreasing phase arrows for log(K/Ca) are smaller than the diamond symbols. The abundances of CI Chondrites \citep[blue square,][]{Lodders2019_solarabund} and Earth's Continental Crust \citep[blue cross-dot,][]{Rumble2019_contcrust} are plotted for comparison. The abundances of main-sequence stars from the SAGA database \citep{Suda2017_SAGAIV} in 1-dex wide [Fe/H] bins centered on 0, –1, –2, and –3 are plotted as solid circles as described in Section \ref{sec:saga_abunds}. We determined the log(Li/Ca) abundances for the main-sequence stars using Equation \ref{eq:infer_saga_LiCa} rather than taking the SAGA database measured log(Li/Ca) abundances because Li abundances from main-sequence stars are underabundant compared to their protostellar nebulae \citep{Fu2015_Limetalpoorstars}. Different hatched regions are included to aid the eye. The values are provided in Table \ref{tab:ssp_abunds}. \label{fig:allel_vs_Ca_ssp}}
\end{figure*}

In the steady-state phase (SSP), accretion and diffusion are in equilibrium. The white dwarf is accreting new planetary material at the same rate at which it is diffusing out of its convection zone. Therefore, in order to recover the planetary material's original abundances, we must adjust them using the relative diffusion timescales. If for example, an element such as Li diffuses more slowly (greater diffusion timescale) from the convection zone than an element such as Ca (lesser diffusion timescale), then in the steady-state phase the amount of Li relative to Ca will be higher in the photosphere than it was in the planetary material. Thus, the inferred Li/Ca abundance of the parent planetary body will be lower than the present photospheric Li/Ca. The equation to recover the abundances of the accreted body if the accretion is in the steady-state phase is given by the following:
\begin{equation}\label{eq:ss_phase}
\log(\text{el}_1/\text{el}_2)_{\text{SSP}}=\log(\text{el}_1/\text{el}_2)_{\text{Phot.}}+\log(\tau_{\text{el}_2}/\tau_{\text{el}_1}),
\end{equation}
where $\log(\text{el}_1/\text{el}_2)_{\text{SSP}}$ is the abundance of the accreted body, and $\tau_{\text{el}}$ is the diffusion timescale (in years) of a given element (see Section \ref{sec:diff_timescales}) \citep{Harrison2018_geoexomaterial}. The inferred abundances of the accreted bodies assuming accretion is in the steady-state phase are provided for each white dwarf in Table \ref{tab:ssp_abunds} and plotted as diamonds in Figure \ref{fig:allel_vs_Ca_ssp}. The reported error bars for the steady-state phase result from a Monte Carlo (MC) using the photospheric relative abundance uncertainties and diffusion timescale uncertainties (see Section \ref{sec:diff_timescales}).

\begin{table}[!ht]
    \caption{Inferred abundances of the accreted planetesimals assuming accretion is in the steady-state phase. We have included the abundances of CI Chondrites, continental crust, and main-sequence stars in 4 metallicity bins for comparison.}
    \centering
    \begin{tabular}{|l|l|l|l|l|l|l|}
    \hline
        Object & $\log(\text{Li}/\text{Ca})_{\text{SSP}}$ & $\log(\text{Na}/\text{Ca})_{\text{SSP}}$ & $\log(\text{Mg}/\text{Ca})_{\text{SSP}}$ & $\log(\text{K}/\text{Ca})_{\text{SSP}}$ & $\log(\text{Cr}/\text{Ca})_{\text{SSP}}$ & $\log(\text{Fe}/\text{Ca})_{\text{SSP}}$  \\ \hline
        WD~J1644$-$0449 & $-2.25\pm0.27$ & $-0.16\pm0.26$ &  & $-1.40\pm0.28$ &  &   \\ \hline
        SDSS~J1330+6435 & $-2.08\pm0.42$ & $0.06\pm0.47$ &  & $<-0.3$ &  &   \\ \hline
        WD~J1824+1213 & $-2.56\pm0.30$ & $-0.62\pm0.31$ & $0.57\pm0.41$ & $<-1.6$ &  &   \\ \hline
        WD~J2317+1830 & $-0.82\pm0.27$ & $0.52\pm0.26$ &  & $<0.9$ &  &   \\ \hline
        LHS~2534 & $-1.93\pm0.25$ & $0.44\pm0.24$ &  & $0.21\pm0.25$ & $0.02\pm0.24$ & $1.22\pm0.30$  \\ \hline
        %WD~J2356$-$209 & $<-2.86$ & $0.86\pm0.28$ & $1.16\pm0.30$ & $<-0.95$ & $<-0.33$ & $1.00\pm0.30$  \\ \hline
        SDSS~J1636+1619 & $<-2.4$ & $<-0.6$ & $<1.0$ & $<-0.4$ &  & $<1.4$ \\ \hline
        \hline
        CI Chondrite\tablenotemark{a} & $-3.03$ & $-0.01$ & $1.24$ & $-1.21$ & $-0.64$ & $1.17$\\\hline
        Continental Crust\tablenotemark{b} & $-2.56$ & $-0.00$ & $-0.03$ & $-0.29$ & $-2.72$ & $-0.01$\\\hline
        \hline
        [Fe/H]$=0.0$\tablenotemark{c} & $-3.48\pm0.17$\tablenotemark{d} & $-0.07\pm0.08$ & $1.25\pm0.08$ & ~ & $-0.65\pm0.07$ & $1.11\pm0.06$  \\ \hline
        [Fe/H]$=-1.0$\tablenotemark{c} & $-2.95\pm0.25$\tablenotemark{d} & $-0.19\pm0.15$ & $1.30\pm0.07$ & ~& $-0.80\pm0.13$ & $0.98 \pm0.09$ \\ \hline
        [Fe/H]$=-2.0$\tablenotemark{c} & $-1.77\pm0.35$\tablenotemark{d} & $-0.39\pm0.30$ & $1.17\pm0.13$ & ~& $-0.99\pm0.25$ & $0.83\pm0.16$ \\ \hline
        [Fe/H]$=-3.0$\tablenotemark{c} & $-0.90\pm0.36$\tablenotemark{d} & $-0.26\pm0.62$ & $1.27\pm0.28$ & ~ & $-1.16\pm0.23$ & $0.86\pm0.27$ \\ \hline
    \end{tabular}
    \raggedright
    \tablenotetext{a}{\citet{Lodders2019_solarabund}}
    \tablenotetext{b}{\citet{Rumble2019_contcrust}}
    \tablenotetext{c}{\citet{Suda2017_SAGAIV} converted from solar-normalized abundances using \citet{Lodders2019_solarabund}.}
    \tablenotetext{d}{Calculated assuming BBN A(Li) from \citet{Coc2014_BBNALi}, [Ca/H] distribution from \citet{Suda2017_SAGAIV}, and $\text{A(Ca)}_{\odot}$ from \citet{Lodders2019_solarabund}. See Equation \ref{eq:infer_saga_LiCa} for calculation.}

    \label{tab:ssp_abunds}
\end{table}

\subsubsection{Decreasing Phase}\label{sec:decreasing}
In the decreasing phase (DP), diffusion dominates. More material is diffusing out of the convection zone than is being newly accreted because the accretion has decreased or halted completely. If the accretion is in the decreasing phase, then the accreted body abundances are given by the following equation:
\begin{equation}\label{eq:dec_phase}
%\log(\text{el}_1/\text{el}_2)_{\text{DP}}=\log(\text{el}_1/\text{el}_2)_{\text{Phot.}}+\log(\tau_{\text{el}_2}/\tau_{\text{el}_1})+t\log(\mathrm{e})*\Bigg(\frac{\tau_{\text{el}_2}-\tau_{\text{el}_1}}{\tau_{\text{el}_2}\tau_{\text{el}_1}}\Bigg),
\log(\text{el}_1/\text{el}_2)_{\text{DP}}=\log(\text{el}_1/\text{el}_2)_{\text{Phot.}}+\log(\tau_{\text{el}_2}/\tau_{\text{el}_1})+t/\ln(10)*\Bigg(\frac{\tau_{\text{el}_2}-\tau_{\text{el}_1}}{\tau_{\text{el}_2}\tau_{\text{el}_1}}\Bigg),
\end{equation}
where $\log(\text{el}_1/\text{el}_2)_{\text{DP}}$ is the inferred accreted body abundance, $t$ is the time since accretion ceased in the same units as $\tau$ \citep{Harrison2018_geoexomaterial}. Note that Equation \ref{eq:dec_phase} is for the scenario in which the accretion was in the steady-state phase just before accretion ceased. If accretion ceased prior to reaching the steady-state phase (i.e. in the increasing phase), then the equation would be the same except the $\log(\tau_{\text{el}_2}/\tau_{\text{el}_1})$ term would be removed.

In the decreasing phase, the photospheric abundances further diverge from the original abundances of the accreted body. Elements with shorter diffusion timescales, such as calcium, deplete from the convection zone more quickly than elements with longer diffusion timescales, such as lithium, resulting in an apparent enhancement in the photospheric abundances of longer-diffusion timescale elements relative to shorter-diffusion timescale elements. For example, photospheric log(Li/Ca) becomes greater the further into decreasing phase accretion progresses, so the accreted body log(Li/Ca) would be lower than the measured photospheric log(Li/Ca) because a disproportionate amount of calcium would have diffused out of the convection zone compared to lithium.

At the same time, the overall abundance of all polluting elements will be decreasing as they are all still diffusing out of the convection zone. Eventually the pollutant abundances will fall below detection thresholds, and the white dwarf will no longer appear to be polluted by accreted planetesimals. Therefore, there is some limit on how far into decreasing phase we could reasonably be catching accretion. We follow \citet{Swan2019_exoplanets} and limit our decreasing phase analysis to 5 calcium diffusion timescales ($\tau_{\text{Ca}}$). We do not show uncertainties in the decreasing phase because it is itself an uncertain phase. It is unclear how much time has passed since accretion ceased, so we plot the potential decreasing phase abundances from 0 to 5~$\tau_{\text{Ca}}$ as faded arrows extending from the steady-state abundances in Figure \ref{fig:allel_vs_Ca_ssp}. The decreasing phase as represented should be considered a general indicator rather than a firm value.

\subsubsection{Diffusion Timescales}\label{sec:diff_timescales}

We use the same method to determine diffusion timescales as was used in \citet{Kaiser2021_LiWD} but updated to use the diffusion timescale tables of \citet{Koester2020_WDenvelope}. Unfortunately our white dwarf \Teff\ values fall below the diffusion timescale grids of \citet{Koester2020_WDenvelope}, so we must extrapolate. We first use the individual diffusion timescale grids to make relative diffusion timescale grids. We then linearly extrapolate the relative diffusion timescale grids to lower \Teff\ values by fitting a line to the log of all \Teff~$<$~10,000 K for a given log($g$) in the grid. We then linearly interpolate between these extrapolated values to obtain relative diffusion timescales for different log($g$) values. We use the diffusion timescales of \citet{Koester2020_WDenvelope} for DB white dwarfs with overshoot=1.0 as \citet{Cunningham2022_xrayg2938planetaccrete} demonstrated convective overshoot is likely present in lower-\Teff, debris-accreting white dwarfs.
Following \citet{Kaiser2021_LiWD}, we also add an additional uncertainty of 0.2 in the relative diffusion timescales by performing a single random-draw Monte Carlo analysis on each diffusion timescale resulting from the \Teff\ and log($g$) Monte Carlo. This additional 0.2 uncertainty is intended to simulate the overall uncertainty resulting from our underlying extrapolation technique compared to a calculation of the diffusion timescales for each individual white dwarf envelope. This additional uncertainty is the dominant source of uncertainty in the relative diffusion timescales.

\begin{table}[!ht]
    \caption{Relative diffusion timescales for the white dwarfs discussed in this work as described in Section \ref{sec:diff_timescales}.}
    \centering
    \begin{tabular}{|l|l|l|l|l|l|l|}
    \hline
        Object & $\log(\tau_{\text{Li}}/\tau_{\text{Ca}})$ & $\log(\tau_{\text{Na}}/\tau_{\text{Ca}})$ & $\log(\tau_{\text{Mg}}/\tau_{\text{Ca}})$ & $\log(\tau_{\text{K}}/\tau_{\text{Ca}})$ & $\log(\tau_{\text{Cr}}/\tau_{\text{Ca}})$ & $\log(\tau_{\text{Fe}}/\tau_{\text{Ca}})$  \\ \hline
        WD J1644$-$0449 & $0.55\pm0.20$ & $0.24\pm0.20$ &  & $-0.01\pm0.20$ &  &   \\ \hline
        SDSS J1330+6435 & $0.58\pm0.20$ & $0.24\pm0.20$ &  & $-0.01\pm0.20$ &  &   \\ \hline
        WD J1824+1213 & $0.51\pm0.20$ & $0.24\pm0.20$ & $0.25\pm0.20$ & $-0.01\pm0.20$ &  &   \\ \hline
        WD J2317+1830 & $0.65\pm0.20$ & $0.25\pm0.20$ &  & $0.00\pm0.20$ &  &   \\ \hline
        LHS 2534 & $0.52\pm0.20$ & $0.22\pm0.20$ &  & $-0.01\pm0.20$ & $-0.16\pm0.20$ & $-0.18\pm0.20$  \\ \hline
        SDSS J1636+1619 & $0.56\pm0.20$ & $0.23\pm0.20$ & $0.24\pm0.20$ & $0.00\pm0.20$ &  & $-0.20\pm0.20$ \\ \hline
    \end{tabular}
    \label{tab:diff_timescales}
\end{table}

\subsection{White Dwarf Ages}\label{sec:WD_ages}
To fully explore the potential explanations of the origin of the Li excess, we must understand several other traits of the white dwarfs in addition to the measured abundances of the planetesimals. The first additional system trait we are interested in is the total age of the white dwarf host and therefore total age of the system because metals were formed over time, so the abundances of a protoplanetary nebula were influenced by the epoch of formation.

\begin{figure*}[htb!]
\includegraphics[scale=1]{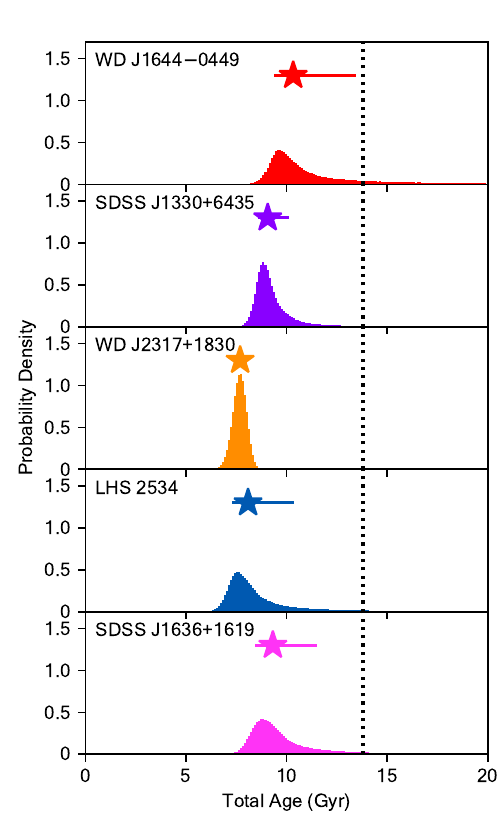}

\centering
\caption{Total Age Distributions of Low Temperature DZs discussed in this work. The star markers are placed at the median and error bars run from the 16th to 84th percentile. The dotted line is at the age of the Universe, 13.8 Gyr \citep{Planck2016_ageofuniverse}. WD~J1824+1213 is not plotted because its mass is too low to yield a viable MC sample of total ages. \label{fig:tot_age_hist}}
\end{figure*}

We compute the total ages of the white dwarf systems using a similar method to that employed in \citet{Kaiser2021_LiWD} with a few minor adjustments. We consider the total age to be the sum of the white dwarf cooling age \citep{Bedard2020_specevoI}, the time for the progenitor to reach the base of the giant branch ($t_{\text{BGB}}$) from \citet{Hurley2000_stellarevo}, and the time of core He burning ($t_{\text{He}}$) from \citet{Hurley2000_stellarevo}. We infer the progenitor mass by inverting the Initial-to-Final Mass Relation (IFMR) of \citet{Cummings2018_IFMR} for each white dwarf mass. 
%so to approximate the age uncertainty introduced by the uncertainty of the hydrogen mass, we assume an 80\% and 20\% probability of $\text{M}_{\text{H}}=10^{-4} \text{M}_\odot$ and $\text{M}_{\text{H}}=10^{-10} \text{M}_\odot$, respectively (\textbf{***citation needed on fractions of each type}). This is achieved by generating a Monte Carlo (MC) sample of $4*10^6$ white dwarf masses and effective temperatures with the $\text{M}_{\text{H}}=10^{-4} \text{M}_\odot$ cooling models using the atmospheric effective temperature and surface gravity uncertainties.
%A MC sample of $1\times10^6$ white dwarf masses and effective temperatures with the $\text{M}_{\text{H}}=10^{-10} \text{M}_\odot$ cooling models.
We generate a Monte Carlo (MC) sample of white dwarf masses and effective temperatures using the \citet{Bedard2020_specevoI} cooling models using the atmospheric effective temperature and surface gravity uncertainties from the model fits in Section \ref{sec:atm_modeling}.
%Unfortunately, we cannot conclusively determine the hydrogen mass of the white dwarfs considered here (see Section \ref{sec:H-layer}), so we allocate a proportional number of MC white dwarfs for each hydrogen mass category as found by \citet{Cunningham2020_Hfractions}. \citet{Cunningham2020_Hfractions} found that 15\% of white dwarfs have $\text{M}_{\text{H}}/\text{M}_{\text{WD}}<10^{-14}$, 25\% of white dwarfs have $10^{-14} \leq \text{M}_{\text{H}}/\text{M}_{\text{WD}}\leq10^{-10}$, and 60\% have $\text{M}_{\text{H}}/\text{M}_{\text{WD}}>10^{-10}$. We believe our white dwarfs most likely have $\text{M}_{\text{H}}/\text{M}_{\text{WD}}>10^{-14}$ as they all have indications of the presence of hydrogen. We therefore allocate simulated white dwarf atmospheres based on the two remaining percentages with $\text{M}_{\text{H}}/\text{M}_{\text{WD}}>10^{-14}$.
For each object, we initialize $2.5\times10^6$ simulated white dwarfs with a ``thin'' H-layer ($\text{M}_{\text{H}}=10^{-10} \text{M}_\odot$) from the cooling models of \citet{Bedard2020_specevoI}. See Section 
\ref{sec:H-layer} for a discussion of our decision to use the ``thin'' H-layer models for all white dwarfs.
The MC white dwarf sample is passed through a MC sample of the IFMR \citep{Cummings2018_IFMR} with its own reported uncertainties. The resulting distribution of progenitor masses was evaluated using the polynomials of \citet{Hurley2000_stellarevo} to determine $t_{\text{BGB}}$ and $t_{\text{He}}$.
We discard simulated white dwarfs for which $\text{M}_{\text{WD}} < 0.532 \text{M}_{\odot}$ or $\text{M}_{\text{WD}} > 1.24 \text{M}_{\odot}$ as the boundaries of the IFMR, and we also discard white dwarfs for which $\text{M}_{\text{WD}}> \text{M}_{\text{progenitor}}$, as a star should not gain mass as it becomes a white dwarf. These overly massive simulated white dwarfs are a result of the uncertainties in the IFMR, and they generally occur for lower masses. We also discard simulated white dwarfs whose total age is $>$20 Gyr to reduce bias near the age of the Universe while preventing clearly non-physical total ages from dragging the distribution to outlandish ages.
The remaining simulated white dwarfs have their respective cooling ages, $t_{\text{BGB}}$, and $t_{\text{He}}$ summed to produce total ages. The resulting total age distributions of the white dwarfs are shown in Figure \ref{fig:tot_age_hist}. We use the median total age as the central value for plots and the 16th and 84th percentiles as the error bars. These values are tabulated in Table \ref{tab:tot_ages}. WD~J1824+1213, however, has an exceptionally low mass ($\text{M}_{\text{WD}}=0.33\pm0.04\text{M}_{\odot}$), so too few simulated white dwarfs made it through the MC to determine an age. Thus, we do not determine a total age for WD~J1824+1213 using this method.

\begin{deluxetable*}{lrr}
\tablecaption{System Total Ages and Progenitor Masses. The median total ages of each white dwarf with the 16th and 84th percentile bounds from the MC sampling as error bars and the corresponding median progenitor masses with 16th and 84th percentile bounds. Only the progenitor masses from the MC sampling with retained total ages are included.
WD~J1824+1213 does not have a total age or progenitor mass listed because its low mass caused the MC sampling to fall outside the prescribed boundaries in Section \ref{sec:WD_ages}. \label{tab:tot_ages}}
\tablewidth{0pt}
\tablehead{
\colhead{Object} & \colhead{Total Age} &\colhead{Progenitor Mass} \\
\colhead{} & \colhead{(Gyr)} & \colhead{($M_{\odot}$)}
}
%\decimalcolnumbers
\startdata
WD~J1644$-$0449 & $10.3^{+3.1}_{-0.9}$& $1.9^{+1.2}_{-0.6}$ \\
SDSS~J1330+6435 & $9.1^{+1.1}_{-0.5}$& $2.8^{+1.7}_{-1.1}$ \\
WD~J1824+1213 & N/A & N/A \\
WD~J2317+1830 & $7.7^{+0.3}_{-0.4}$ & $5.4^{+1.3}_{-1.0}$ \\
LHS~2534 & $8.1^{+2.3}_{-0.8}$ & $2.0^{+0.9}_{-0.6}$ \\
%WD~J2356$-$209 & $12.3^{+3.6}_{-1.6}$ \\
SDSS~J1636+1619 & $9.3^{+2.2}_{-0.9}$& $1.9^{+0.8}_{-0.6}$ \\
\enddata
\end{deluxetable*}

\subsection{Galactic Kinematics}\label{sec:kinematics}
\begin{figure*}[htb!]
\includegraphics[scale=1]{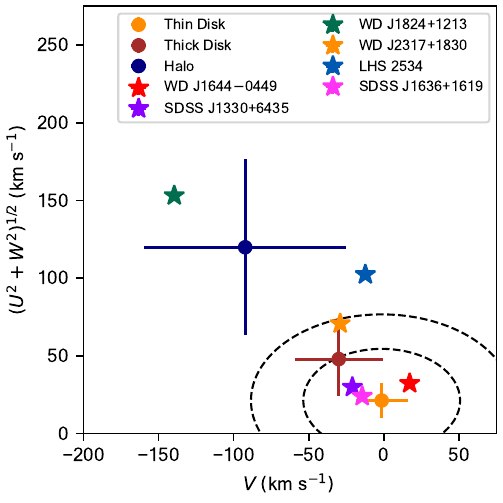}
\centering
\caption{Toomre Diagram showing peculiar velocities relative to the local standard of rest. The white dwarfs discussed in this work are shown as star symbols. The errorbars of the white dwarfs' velocities are smaller than the symbols. The mean velocities and dispersions of the halo, thick disk, and thin disk are shown are shown as the blue, maroon, and orange circles, respectively \citep{Torres2019_WDKinematics}. The 3$\sigma$ and 5$\sigma$ dispersion ellipses of the thin disk are shown as dashed lines.   \label{fig:toomre}}
\end{figure*}

The kinematic membership of a given system also correlates with the relative abundances of elements in the protoplanetary nebula \citep[e.g.][]{Bensby2014_generalabunds}.
We computed the Cartesian peculiar velocities ($U,V,W$) relative to the local standard of rest (LSR) for the white dwarfs discussed in this work using parallaxes and proper motions from {\it Gaia} DR3 \citep{Gaia2022_GaiaDR3Summary} and the Astropy coordinates package \citep{astropy2013,astropy2018,astropy2022}. We set the radial velocities to zero following previous white dwarf kinematic studies lacking radial velocities \citep{Kilic2019_HaloAge, Torres2019_WDKinematics, Kaiser2021_LiWD}. Our spectroscopy is not adequate to derive radial velocities.
%The high-pressure atmospheres of low temperature white dwarfs can shift the wavelengths of metal lines in addition to broadening \citep[see Figure 1 of][]{Blouin2018_PaperI}. Additionally, our spectroscopy was generally low resolution, and the one instance of higher resolution spectroscopy targeted the strongly magnetic LHS~2534. Magnetism also shifts metal absorption lines and can lead to asymmetries, further confounding potential radial velocity measurements \citep{HollandsStopkowicz2023_DZHmagneticmodeling}.
$U$ is the velocity along the axis pointing toward Galactic center. $V$ is the velocity in the direction of Galactic rotation in the plane of the Galaxy, so $V<0$ indicates motion lagging the Galactic rotation. $W$ is the velocity normal to the Galactic plane. We provide the $U,V,W$ velocities in Table \ref{tab:kinematics}, and we plot them in a Toomre diagram in Figure \ref{fig:toomre}.
\begin{deluxetable*}{lrrcc}
\tablecaption{Peculiar Velocities of White Dwarfs and Galactic Kinematic Memberships. \label{tab:kinematics}}
\tablewidth{0pt}
\tablehead{
\colhead{Object}  & \colhead{$V$} & \colhead{$(U^2+W^2)^{(1/2)}$} & \colhead{Membership} & \colhead{\citet{Torres2019_WDKinematics}}\\
\colhead{} & \colhead{(km/s)} & \colhead{(km/s)} & \nocolhead{Name} & \colhead{}
}
%\decimalcolnumbers
\startdata
WD~J1644$-$0449 & $17.2\pm0.5$ & $32.3\pm2.3$ & Thin or Thick Disk & ~ \\
SDSS~J1330+6435 & $-21.0\pm2.2$& $29.9\pm2.1$ & Thin or Thick Disk & ~ \\
WD~J1824+1213 & $-139.5\pm1.1$ & $153.2\pm1.1$ & Halo & Halo \\
WD~J2317+1830 & $-29.1\pm0.5$ & $70.8\pm0.8$& Thick Disk or Halo & Thick Disk\\
LHS~2534 & $-12.3\pm0.1$& $102.4\pm0.4$ & Thick Disk or Halo & Thick Disk \\
%WD~J2356$-$209 & $-13.2\pm1.0$&$124.1\pm4.4$& Thick Disk or Halo & Thick Disk \\
SDSS~J1636+1619 &$14.4\pm0.6$& $24.1\pm0.3$& Thin or Thick Disk & Thin Disk\\ \hline
Thin Disk\tablenotemark{a} & $-1.5\pm 17.4$ & $21.2\pm11.1$& ~ & ~ \\
Thick Disk\tablenotemark{a} & $-30.0\pm29.3$ & $47.8\pm23.6$ & ~ & ~ \\
Halo\tablenotemark{a} & $-92.3\pm67.4$ & $119.8\pm56.3$ & ~ & ~\\
\enddata
%\tablerefs{\citep{Torres2019_WDKinematics}}
\tablenotetext{a}{Values from \citet{Torres2019_WDKinematics}.}
\end{deluxetable*}

If we compare the $U,V,W$ velocities of our white dwarfs with mean velocities and dispersions of the three Galactic kinematic populations (halo, thick disk, and thin disk) from \citet{Torres2019_WDKinematics}, we can constrain their membership. WD~J1644–0449, SDSS~J1330+6435, and SDSS~J1636+1619 all fall within the 3$\sigma$ ellipse of the thin disk, so they are most likely in the thin disk, but it is possible they are actually in the thick disk given the overlap in the two populations. WD~J2317+1830 is the most ambiguous in its membership. WD~J2317+1830 falls between the 3 and 5$\sigma$ ellipses for the thin disk, so it is most likely not in the thin disk. Its velocity is compatible with the thick disk, but it is also compatible with the halo. Therefore, we place it in the thick disk or halo. LHS~2534 has confounding kinematics as well. Its position outside the 5$\sigma$ ellipse in the Toomre diagram (Figure \ref{fig:toomre}) rules out thin disk membership and indicates halo, but its near co-rotation with the disk ($V\approx0$) suggests thick disk \citep{Chiba2000_halorotvelocity}. We therefore assign LHS~2534 to the thick disk or halo. 
%\citet{Torres2019_WDKinematics} also assigned kinematic populations for several white dwarfs discussed in this work using Gaia DR2 astrometry \citep{Lindegren2018_GaiaDR2astrometry,Gaia2018_DR2summary}. 
WD~J2317+1830 and LHS~2534 were assigned to the thick disk by \citet{Torres2019_WDKinematics} in agreement with our findings. %\citet{Torres2019_WDKinematics} also concurred that WD~J1824+1213 is a member of the halo.
WD~J1824+1213 has a peculiar velocity that places it firmly in the halo as previously determined by others \citep{Torres2019_WDKinematics, Hollands2021_LiWD,Elms2022_newLiWD, Bergeron2022_ultracoolWDsarenotcool}. Its lack of Galactic co-rotation ($V$) is typical of halo membership in addition to its elevated $(U^2+W^2)^{1/2}$.
The only thin disk assignment of \citet{Torres2019_WDKinematics} among our white dwarfs is SDSS~J1636+1619. WD~J1644–0449 and SDSS~J1330+6435 were not included in the catalogue of \citet{Torres2019_WDKinematics}. We include the membership assignments of \citet{Torres2019_WDKinematics} in Table \ref{tab:kinematics}.

\subsection{Expected Galactic Nucleosynthetic Abundances from the SAGA Database}\label{sec:saga_abunds}

\begin{figure*}
\gridline{\fig{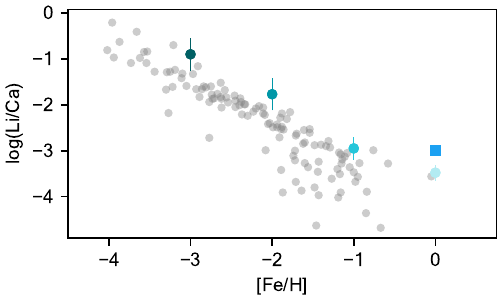}{0.48\textwidth}{(a)}
    \fig{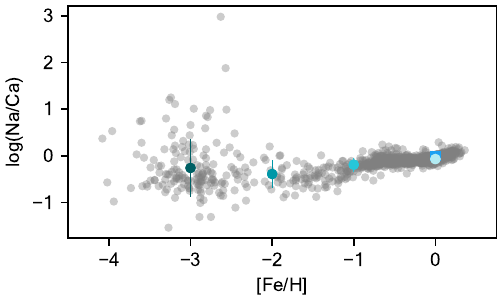}{0.48\textwidth}{(b)}}
\gridline{\fig{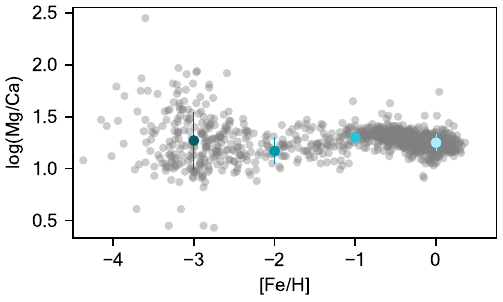}{0.48\textwidth}{(c)}
    \fig{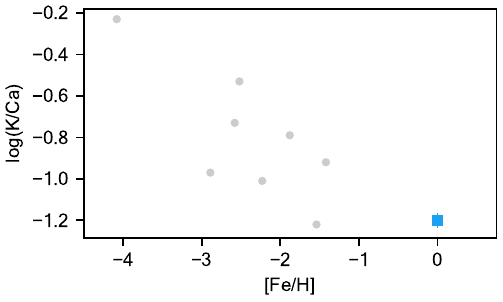}{0.48\textwidth}{(d)}}
\gridline{\fig{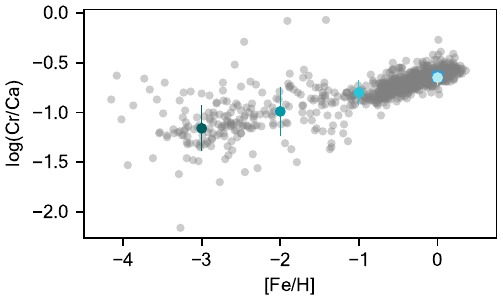}{0.48\textwidth}{(e)}
    \fig{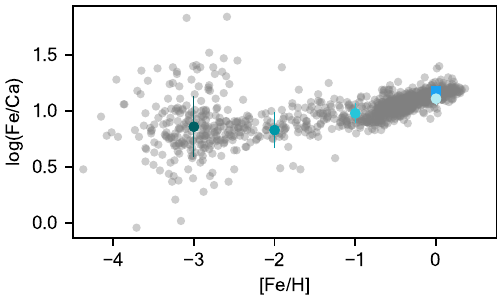}{0.48\textwidth}{(f)}}

\caption{Galactic Nucleosynthetic Evolution as shown by Main-Sequence Stars from the SAGA Database. The [Fe/H] abundances on the x-axis are plotted in the Solar System-normalized format ($[\text{el}/\text{Ca}]=\log(\text{el}/\text{Ca})-\log(\text{el}/\text{Ca})_{\odot})$). The grey circles are main-sequence stars from the SAGA database \citep{Suda2017_SAGAIV}. The circles in various shades of blue represent the mean and standard deviation of the log(el/Ca) abundance of the main-sequence stars in a given 1-dex wide [Fe/H] bin centered on the integer [Fe/H] values as described in Section \ref{sec:saga_abunds}. These are the log(el/Ca) points plotted in Figures \ref{fig:allel_vs_Ca_phot} and \ref{fig:allel_vs_Ca_ssp}. The blue square represents the Solar System abundances from \citet{Lodders2019_solarabund} for a given log(el/Ca). By definition, [Fe/H]=0 for the Solar System. The binned points of Panel (a) do not align with the SAGA datapoints because the binned points for log(Li/Ca) are intended to represent the protostellar nebulae (calculated using Equation \ref{eq:infer_saga_LiCa}), and Li abundances from main-sequence stars are underabundant compared to their protostellar nebulae \citep{Fu2015_Limetalpoorstars}. There are no binned points in Panel (d) because there are so few main-sequence stars in the SAGA database with K, Ca, and Fe measured with uncertainties provided.}
\label{fig:saga_abunds}
\end{figure*}

We wish to examine protoplanetary nebular abundances down to extremely low metallicities ([Fe/H]~$<-2.5$), which requires knowledge of the relative abundances of the other elements being investigated (Li, Na, Mg, K, Ca, and Cr) at those extreme metallicities as well. Given the rarity of extremely-metal-poor (EMP) stars, an aggregation of EMP stars from multiple works is appropriate.
We attempt to provide a representation of the Galactic nucleosynthetic evolution of the elements under consideration via the abundances of ``main-sequence'' (non-RGB) stars from the ``recommended'' abundance list of the Stellar Abundances for Galactic Archaeology (SAGA) Database \citep{Suda2017_SAGAIV}. We use the SAGA database instead of a single dataset such as \citet{Bensby2014_generalabunds} because we need a dataset that extends to extremely low metallicities ([Fe/H]~$<-2.5$) and provides abundances for as many of the elements being investigated as possible for the largest number of stars possible; detailed chemical abundances of multiple elements are published for around two dozen extremely-metal poor stars at a time in the best of cases \citep[e.g.][]{Mardini2019_6EMPstars, Venn2020_PristineIX28EMPstars}. Thus, while there will be some inherent systematics related to using data from multiple sources, the benefit from sampling the extremely low metallicities outweighs the downsides.

To exclude RGB stars, we employ the \Teff~and log($g$) cuts used in the database: \Teff$>6000$~K or $\log(g)>3.5$. We also only include those stars for which the abundances are accompanied by uncertainties, so this should exclude measurements without reported uncertainties and measurements that are actually upper limits. We check all of the ionization states of each metal as well. We then convert the solar-normalized [el/Ca] abundances to the general log(el/Ca) number abundances using the Solar System abundances of \citet{Lodders2019_solarabund}. We then produce representative data points (blue circles in Figure \ref{fig:saga_abunds}) for each 1-dex wide [Fe/H] bin centered on the integer [Fe/H] values by taking the mean and standard deviation of the log(el/Ca) of the stars within a given [Fe/H] bin. The mean and standard deviation of each [Fe/H] bin are included in Table \ref{tab:ssp_abunds}. These representative points are included in Figures \ref{fig:allel_vs_Ca_phot} and \ref{fig:allel_vs_Ca_ssp} to visualize Galactic nucleosynthetic evolution. There are too few main-sequence stars in the database to perform this process for log(K/Ca), so we do not attempt to create representative points for log(K/Ca).

Main-sequence stars are also known to deplete their Li compared to their protostellar nebulae (e.g. \citet{Prantzos2012_Gal_LiBeB,Fu2015_Limetalpoorstars}), so the log(Li/Ca) that would be directly derived from the SAGA database (grey circles in Figure \ref{fig:saga_abunds} panel (a)), would be an underestimate of the log(Li/Ca) of the protoplanetary nebulae that formed the planetesimals accreted by the white dwarfs in this work. Therefore, we instead estimated the protostellar (and therefore protoplanetary) nebular log(Li/Ca) abundances of each star in the SAGA database using the Li abundance from the Big Bang and the Ca abundance from each SAGA star. This relation is given by Equation \ref{eq:infer_saga_LiCa}:
\begin{equation}\label{eq:infer_saga_LiCa}
    \log(\text{Li/Ca})=\text{A(Li)}_{\text{BBN}}-\text{[Ca/H]}-\text{A(Ca)}_{\odot},
\end{equation}
where $\text{A(Li)}_{\text{BBN}}$ is the Big Bang nucleosynthetic Li abundance from \citet{Coc2014_BBNALi}\footnote{$\text{A(el)}\equiv \text{log(el/H)}+12$}, $\text{[Ca/H]}$ is the solar-normalized abundance for each star from the SAGA database\footnote{$\text{[el/H]}\equiv\text{log(el/H)}-\text{log(el/H)}_{\odot}$}, and $\text{A(Ca)}_{\odot}$ is the Solar System abundance from \citet{Lodders2019_solarabund}, which is necessary to convert the star's solar-normalized Ca abundance to the A(el) convention. The depletion of Li in main-sequence stars is the reason the binned data points lie above the main-sequence star points in Panel (a) of Figure \ref{fig:saga_abunds}. Unfortunately, this method still leads to an underestimate of the protoplanetary log(Li/Ca) as it does not account for ongoing synthesis of Li. The underestimate is initially small at the lowest metallicities, but approaches $\approx0.5$~dex as demonstrated by our log(Li/Ca) estimate for solar metallicity falling short of the CI Chondrite log(Li/Ca) (See the blue square in Panel (a) of Figure \ref{fig:saga_abunds}).

\section{Spallation}\label{sec:spallation}
Before we evaluate the other two hypotheses, geologic differentiation and Big Bang combined with Galactic nucleosynthesis,
we first consider the spallation hypothesis and find it is not feasible to explain the Li excesses in these 5 white dwarfs.
\citet{Klein2021_BeWD} discovered that two white dwarfs were polluted by planetary material enhanced in beryllium by $\sim$2 dex relative to CI Chondrites: GALEX 2667197548689621056 (hereafter GALEX J2339$-$0424) and GD 378. Noting the high-energy radiation environment of Jupiter \citep{Jun2005_jupiterprotons} and its and Saturn's rings and moons, \citet{Doyle2021_spallWD} hypothesized that this Be enhancement originated from the accretion of an icy exomoon formed from icy rings orbiting a gas giant planet (or brown dwarf). This icy ring material was spalled by protons originating from the white dwarf progenitor's stellar wind that were trapped in the exoplanet's magnetic field. The Be-enhanced rings then coalesced to form an icy exomoon that the white dwarf eventually accreted resulting in the measured photospheric abundances. \citet{Doyle2021_spallWD} pointed out that a corollary of this explanation is that a certain level of Li enrichment would also be expected as another nucleosynthetic source of Li is spallation \citep{Prantzos2012_Gal_LiBeB}. 
%The only nucleosynthetic source of Be is spallation \citep{Prantzos2012_Gal_LiBeB}.
Therefore, the icy exomoon hypothesis has no Galactic epoch requirement because the nucleosynthesis (via spallation) occurs in the radiation belts of a gas giant (or brown dwarf), so it should occur at some consistent frequency regardless of total system age.

There are two issues though with the assumptions used by \citet{Doyle2021_spallWD}: threshold energies and proton energies of gas giant planet radiation belts. Spallation reactions have what are known as threshold energies, minimum energies for the incident proton (or alpha particle) below which the reaction will not occur (i.e. cross-section is 0). For the proton + oxygen-16 yielding beryllium-9 reaction, the threshold energy is 33.7 MeV \citep{Read1984_nukecross}. That means any proton in the radiation belts with an energy less than 33.7 MeV will not be able to produce beryllium by hitting an oxygen-16 atom. Therefore, the peak Jovian flux $\sim10^{7}$ $\text{cm}^{-2}$ $\text{s}^{-1}$ of protons with energies $\geq$10 MeV used by \citet{Doyle2021_spallWD} is an overestimate. We can produce a slightly better estimate by taking the flux of protons with energy $\geq$33.7 MeV from the radiation belt models of \citet{Jun2005_jupiterprotons} for Europa's orbital distance and then rescaling to the higher radiation environment. The result is a proton flux of $\approx4\times10^{5}$ $\text{cm}^{-2}$ $\text{s}^{-1}$ for protons with energy $\geq$33.7 MeV in the highest proton flux environment around Jupiter. Furthermore, the $\sim$10 mb cross section used by \citet{Doyle2021_spallWD} is the maximum cross-section of this nuclear reaction, and it is the cross-section for a proton energy of $\approx$~65 MeV. Even at proton energies of 50 MeV the cross-section is only $\approx$~4 mb. If we estimate the proton flux of Jupiter's high-radiation zone for protons with energy $\geq$50 MeV, it is only $\approx3\times10^4$ $\text{cm}^{-2}$ $\text{s}^{-1}$. The proton flux used by \citet{Doyle2021_spallWD} was more than 300 times this level. If we use this $\geq$50 MeV proton flux with the 10 mb cross-section (an overestimate of the cross-section for this energy range, meaning the reaction is treated as more efficient than it actually is) the radiation timescale for the production of observed beryllium would be on the order of $10^9$ yr. The true timescale is likely longer given that we used an upper limit on the cross section.

This is problematic for the exomoon hypothesis to explain the Li excess in WD~J2317+1830. The progenitor of WD~J2317+1830 likely only survived for $\sim10^8$ yr at most based on the high mass of WD~J2317+1830 and the IFMR of \citet{Cummings2018_IFMR}.\footnote{Given the high mass of WD~J2317+1830, it is possible that it is a merger product, in which case the merger progenitors would have had lesser masses and therefore main-sequence lifetimes that could have been greater than 1 Gyr. However, this would also mean the total age of the system would have to be even greater (the cooling age already accounts for $\sim7.5$~Gyr of the total age), which would put it further into the Galactic epoch where Big Bang and Galactic nucleosynthesis would be likely to have already led to an elevated level of lithium. See Section \ref{sec:nuke_vs_geo} for an explanation of the effect of large total ages on abundances.}
%Therefore for the Li excess of WD~J2317+1830 to originate from this mechanism, a Saturn-like ring system would have to exist around a Jupiter-like planet for the entire life of the progenitor star, and both the ring and progenitor star would have to survive 10 times longer than the progenitor star is presently estimated to live.
Therefore for the Li excess of WD~J2317+1830 to originate from this mechanism, a Saturn-like ring system would have to exist around a Jupiter-like planet that is subjected to stellar wind for 10 times longer than the lifetime of the progenitor star.
This ring material would then have to coalesce to form a moon that does not contain any significant portion of non-spalled material because non-spalled material would dilute the spalled material and lower the overall $\log(\text{Li}/\text{Ca})$ and $\log(\text{Be}/\text{Ca})$ abundances. The spalled material cannot just be accreted as outer layers onto an existing moon; it must be the dominant portion of the mass to yield the observed abundances. Finally, this specific spalled-material moon must be accreted by the white dwarf. \citet{Trierweiler2022_exomoonfrequency} recently estimated the fraction of white dwarfs polluted by exomoons (of any kind) to be about one percent, so of that one percent of pollution events, we also have the even smaller percent of exomoons that are comprised of heavily spalled former ring material. No moons in the Solar System are known to be formed of heavily spalled former ring material. As this requires a series of flukes and seemingly impossible timescales, we rule out the exomoon hypothesis to explain the Li excess in WD~J2317+1830 specifically, and we find it unlikely to explain the Li excess in the other white dwarfs given the low probability of accretion of any exomoon.

%\subsection{Discussion of Individual Cool DZs with Lithium Detections}\label{sec:ind_WDs_wLi}
\section{Big Bang and Galactic Nucleosynthesis versus Geologic Differentiation}\label{sec:nuke_vs_geo}

We now assess some differences between the predictions of the two remaining hypotheses for the Li excesses: a combination of Big Bang and Galactic nucleosynthesis versus geologic differentiation.
In the first model, excess Li/Ca arises as a consequence of the initial abundance of the nebula for very old planetary systems; the nebulae have Li from the Big Bang but have not yet been enriched in Ca.
In the second hypothesis, the Li excess comes from alteration of the nebular abundances by differentiation in planets. But in both hypotheses, we must consider thermal alterations as well, which can happen in planet formation or subsequent heating (e.g. by the central star in its red giant phase), and results in a greater loss of the more volatile elements relative to the more refractory elements.
%Before doing so, however, some additional background knowledge is needed. We introduce thermal alteration as a potential adjustment to abundances that is not unique to either hypothesis as many rocky bodies undergo heating in the Solar System, and it is not limited to differentiated bodies.
%Many rocky bodies undergo (or have undergone) heating in the Solar System, and it is not limited to differentiated bodies.
%Otherwise primitive bodies from ancient protoplanetary nebulae could be heated resulting in the loss of volatiles in a manner analogous to the CK Chondrites in the Solar System. It also follows that continental crust could be subjected to heating at some point in its evolution, such as the AGB phase, which would likewise cause volatile loss.
Thermal alteration presents the opportunity to adjust the abundances predicted by each hypothesis, and we invoke it when we are otherwise unable to find an adequate solution in the context of a given hypothesis. 

The primordial lithium abundance for the Solar System is derived from CI Chondrite meteorites \citep{Lodders2019_solarabund}. CI Chondrites are chemically primitive, which means their elemental abundances are consistent with the protosolar nebula. This was determined by their consistency in abundances with the Sun in all elements save for lithium \citep[and the most volatile elements such as noble gases,][]{Lodders2009_abundsreview}. Presumably, all of the planetary material in the Solar System condensed from this same cloud starting from the same abundances. Depending upon the environment of where the planet formed (i.e. distance from the Sun), different elements condensed more efficiently; the result was a condensation-volatilization sequence. Heating of these planetesimals via collisions and radioactive decay caused volatiles to be lost to space \citep{ONeill2008_volatileloss}. Therefore planetesimals subjected to this post-nebular volatilization would be depleted in volatiles. Essentially, heating causes more volatile elements to deplete in a given planetesimal whether it occurs during the formation or post-formation. Therefore, heating should deplete, in order, Na, K, Li, Cr, Fe, Mg, and Ca \citep{Lodders2003_abund}. This is why Ca is a good denominator for our elemental abundance ratios (in addition to its detection in all of the white dwarfs under consideration); heating cannot cause a spurious enhancement in el/Ca ratios because Ca is the least volatile element detected in these white dwarfs. Using Na, for example, could cause an apparent enhancement in lithium abundance if the planetesimal were subjected to heating of sufficient level to deplete Na but not Li because Li/Na would increase as the amount of Na decreased. It is important to keep this limitation in mind as we examine the abundances of the individual accreted planetesimals in the next section.

Sufficiently massive planets (like Earth) also differentiated to form cores and crusts with siderophiles, such as Fe and Cr, settling preferentially in the core and lithophiles, such as Li, Na, K, and Ca, settling preferentially in the crust \citep{Lodders2009_abundsreview}. However, the net effect of differentiation and Earth's formation only yielded elevated Li/Ca and K/Ca in the continental crust compared to CI Chondrites \citep{Rumble2019_contcrust}. The Na/Ca in continental crust ended up at essentially the same abundance as CI Chondrites. \citet{Hollands2021_LiWD} hypothesized that the Li-polluted white dwarfs had accreted exoplanetary continental crust to explain the Li excess, and we evaluate that hypothesis for each accreted planetesimal in light of abundance ratios for the Earth's crust in the next section.

In the Big Bang and Galactic nucleosynthetic model of enhanced Li, differences in the abundances of the protoplanetary nebula should propagate through the formation process to produce planetesimals with different abundances than are found in the Solar System. If for example the Li/Ca abundance of an extrasolar protoplanetary nebula were 1 dex greater than the Li/Ca abundance of the protosolar nebula, this extrasolar system's analogue of CI Chondrites should have a Li/Ca abundance 1 dex greater than the Li/Ca abundance of Solar System CI Chondrites. These nebular abundances are mostly tied to the metallicity ([Fe/H]) of a system, which is a rough proxy for the age of the system \citep{Kobayashi2020_originofelements}. Due to radial migration, among other factors, there is not a direct age-metallicity relation \citep[e.g.][]{Holmberg2007_agemetallicity, Casagrande2011_agemetallicity}. H, He, and some Li were formed via Big Bang nucleosynthesis around 13.8~Gyr ago \citep{Planck2016_ageofuniverse}, but no heavier elements were formed \citep{Coc2014_BBNALi}. Therefore, at the earliest epochs, the Li/Ca (or Li-to-any-other-metal) ratio should have been much higher than it was in the protosolar nebula which condensed to form the planets and Sun only 4.6~Gyr ago \citep{Bouvier2010_solarsystemage} as shown in Figure \ref{fig:saga_abunds}a. Examining the other panels of Figure \ref{fig:saga_abunds}, we can see the evolution of the other elements discussed in this work.
Other elements were produced at slightly different rates via Galactic nucleosynthesis \citep{Kobayashi2020_originofelements}. The Na/Ca ratio appears to decrease with decreasing metallicity, but the scatter becomes large as $[\text{Fe/H}]\rightarrow-3.0$, so it is possible to have super-solar Na/Ca at sufficiently low metallicity. Unfortunately, K/Ca could not be reliably inferred for enough stars in the SAGA database \citep{Suda2017_SAGAIV} to produce a mean K/Ca for each [Fe/H] bin, so our analysis using K is limited. This is likely a consequence of the \ion{K}{1} resonance lines' proximity to an $\text{O}_2$ telluric absorption feature, which makes retrieving K abundances more difficult.
However, \citet{Kobayashi2020_originofelements} aggregated different samples of [K/Fe] and [Ca/Fe] measurements spanning $-4.0\leq\text{[Fe/H]}<0.5$, and the trends of K and Ca remain within 0.2~dex of each other over the entire metallicity range \citep[see Figures 16 and 21 from][]{Kobayashi2020_originofelements}. Therefore we tentatively proceed under the premise that the nebular log(K/Ca) is roughly solar at all metallicities.
The Mg/Ca ratio appears to be constant with metallicity, which follows as they are both $\alpha$~elements. Cr/Ca and Fe/Ca both decrease with decreasing metallicity, but the effect is less than 0.5~dex.
The Big Bang and Galactic nucleosynthetic hypothesis for enhanced Li/Ca is further complicated by the possibility of thermal alteration and differentiation, which can combine with primordial differences.
%To further muddy the situation, a given planetesimal could be subjected to heating, differentiation, and protoplanetary nucleosynthetic abundance differences, or it could be subjected to only some of these processes.

In Sections \ref{sec:J1644}, \ref{sec:J1330}, \ref{sec:J1824}, \ref{sec:J2317}, \ref{sec:LHS2534}, and \ref{sec:SDSSJ1636} we evaluate each observed system in the context of each of these two remaining hypotheses to explain the Li excess. In the first subsection for each white dwarf (e.g. Section \ref{sec:J1644_bbn}), we evaluate the compatibility of the accreted planetesimal abundances (from Section \ref{sec:infer_abund}) with the abundances of a primitive planetesimal from protoplanetary nebulae in the metallicity bins established in Section \ref{sec:saga_abunds}, for which the reader may reference Figure \ref{fig:allel_vs_Ca_ssp} for visual aid. We also consider thermal alteration to the planetesimal to explain deviations of abundances. We then determine if the metallicity bins compatible with the accreted planetesimal abundances are also consistent with the metallicities that could be expected for the ages and kinematic memberships of the systems determined in Sections \ref{sec:WD_ages} and \ref{sec:kinematics}. In the second subsection for each white dwarf (e.g. Section \ref{sec:J1644_geo}), we evaluate the compatibility of the accreted planetesimal abundances with the abundances of continental crust without primordial abundance ratio differences in the nebulae. The reader may again refer to Figure \ref{fig:allel_vs_Ca_ssp} in general for the planetesimal abundance discussion. As the Earth and its continental crust were formed from a solar metallicity protoplanetary nebula, we also evaluate the system age (Section \ref{sec:WD_ages}) and kinematic membership (\ref{sec:kinematics}) for compatibility with solar metallicity. We present a summary of our evaluation of each system's compatibility with continental crust and the combination of Big Bang and Galactic nucleosynthesis in Table \ref{tab:excess_explanations}.

\begin{table}[ht!]
    \caption{Summary of the compatible accretion scenarios for each of the white dwarfs discussed in Section \ref{sec:nuke_vs_geo}. If thermal alteration of the material is required to be compatible, that is indicated. If a system is incompatible with a hypothesis, that is also indicated. See individual subsections in Section \ref{sec:nuke_vs_geo} for each hypothesis for each white dwarf for elaboration on compatible scenarios.
    \label{tab:excess_explanations}}
    \centering
    \begin{tabular}{|l|l|l|}\hline
        System & Big Bang and Gal. Nucleosynthesis & Geologic Differentiation \\ 
        ~&~&~\\\hline
        WD~J1644–0449 &  Primitive planetesimal from & Heavily thermally-altered continental\\
        ~&the $-2.5\leq\text{[Fe/H]}\leq-0.5$ bins& crust in decreasing phase accretion\\
        ~& or a CI Chondrite  & ~\\
        \hline 
        SDSS~J1330+6435\tablenotemark{a} &  Primitive planetesimal from the& Unaltered continental crust\\
        ~& $-3.5\leq\text{[Fe/H]}\leq-0.5$ bins&  ~\\
        ~& or a CI Chondrite & ~\\
        \hline 
        WD~J1824+1213 & Thermally-altered planetesimal from& Halo membership means it cannot have\\
        ~&the $-2.5\leq\text{[Fe/H]}\leq-0.5$ bins& Solar System abundances nor\\ 
        ~&~& Earth-like continental crust\\\hline
        WD~J2317+1830&Primitive planetesimal from the&  log(Li/Ca) too high to be\\
        ~& $-3.5\leq\text{[Fe/H]}\leq-2.5$ bin&normal continental crust \\ \hline
        LHS~2534& log(Cr/Ca) (and log(K/Ca) probably) & log(Fe/Ca) and log(Cr/Ca) too high for \\ 
        ~& too high for Galactic Nucleosynthesis&continental crust\\\hline
        SDSS~J1636+1619&Planetesimal from any metallicity bin & Continental crust that was heavily \\
        ~&that was heavily thermally altered at& thermally-altered at high temperature  \\
        ~& high temperature&~\\\hline
    \end{tabular}
    
    \tablenotetext{a}{\raggedright SDSS~J1330+6435 has such large error bars and so few measured abundances that it is compatible with nearly any scenario. See Section \ref{sec:J1330}.}
\end{table}

\subsection{WD~J1644–0449}\label{sec:J1644}
\subsubsection{WD~J1644–0449 Big Bang and Galactic Nucleosynthesis}\label{sec:J1644_bbn}

The planetesimal accreted by WD~J1644–0449 can be explained by the accretion of a primitive planetesimal from a protoplanetary nebula with a metallicity $-2.5<\text{[Fe/H]}<-0.5$ or an analogue to a Solar System CI Chondrite.
%We turn the reader's attention to Figure \ref{fig:allel_vs_Ca_ssp} for the discussion of relative abundances.
The \LiCaSSP\ of WD~J1644–0449 is within 3$\sigma$ of the log(Li/Ca) of the [Fe/H]~=~-1.0 and –2.0 main-sequence bins and CI Chondrites as shown in Figure \ref{fig:allel_vs_Ca_ssp}. The \NaCaSSP\ of WD~J1644–0449 is also compatible with that of the [Fe/H]~=~-1.0 and –2.0 bins and CI Chondrites. As stated in Section \ref{sec:saga_abunds}, there is not an adequate number of main-sequence stars in the SAGA database \citep{Suda2017_SAGAIV} with simultaneous K, Ca, and Fe measurements to produce binned points for log(K/Ca) as was done with the other elements. However, assuming log(K/Ca) behaves somewhat similarly to the other metals (excluding Li), i.e. remaining roughly the same as the Solar System, the \KCaSSP~of WD~J1644–0449 is compatible with the accreted planetesimal originating from a protoplanetary nebula in the [Fe/H]~=~-1.0 and –2.0 bins and CI Chondrites. The large age of WD~J1644–0449 ($10.3^{+3.1}_{-0.9}$~Gyr), also points to a sub-solar metallicity protoplanetary nebula. The kinematics of WD~J1644–0449 as shown in Figure \ref{fig:toomre}, however, point to thin disk membership, which would be problematic for the [Fe/H]~=~–2.0 bin but not necessarily the -1.0 bin, as \citet{Bensby2014_generalabunds} found no thin disk stars with $\text{[Fe/H]}<-0.7$, but other studies \citep[e.g.][]{Mishenina2004_lowFeHthindisk} claimed a few thin disk stars with $\text{[Fe/H]}\approx-1$. Thin disk membership is fully compatible with a CI Chondrite analogue with Solar System abundances.
%This still would not be a low enough metallicity though assuming the accretion is in the steady-state phase.
%It is also possible WD~J1644–0449 is actually a member of the thick disk given the intermixing of the thick and thin disk in kinematic space in the Toomre diagram \citep{Torres2019_WDKinematics}(Figure \ref{fig:toomre}).
The [Fe/H]~=~–2.0 bin requires that WD~J1644–0449 be a member of the thick disk,\footnote{Alternatively halo membership would reach these low metallicities, but that is unlikely to be the case.} which is possible given the unknown radial velocity and the intermixing of the thick and thin disk in kinematic space in the Toomre diagram \citep{Torres2019_WDKinematics}(Figure \ref{fig:toomre}).
%The unknown radial velocity of WD~J1644–0449 further muddies the potential kinematic membership, so the accretion of a primitive body formed from a sub-solar metallicity protoplanetary nebula is possible. 

\subsubsection{WD~J1644–0449 Geologic Differentiation}\label{sec:J1644_geo}

The kinematics of WD~J1644–0449 suggest a thin disk membership and therefore allow metallicity not too far from solar for the protoplanetary nebula. Thus the planets and planetesimals could have Solar System abundances, and they could have differentiated to form continental crust-like material. 
The \LiCaSSP~of WD~J1644–0449 is within 2$\sigma$ of continental crust as shown in Figure \ref{fig:allel_vs_Ca_ssp}. The \NaCaSSP~of WD~J1644–0449 is also compatible with continental crust.

The \KCaSSP~of WD~J1644–0449, however, spells trouble for continental crust. The \KCaSSP~of WD~J1644–0449 is more than 3$\sigma$ below the continental crust log(K/Ca). K has a diffusion timescale that is essentially the same as that of Ca \citep{Koester2020_WDenvelope}. Therefore, even if the accretion is in decreasing phase, log(K/Ca) does not change.
The only way to make the abundances compatible with the continental crust hypothesis is to assume thermal alteration.
When heated, log(Na/Ca) and log(K/Ca) deplete at roughly the same rate \citep[see Figure 2 of][]{Kaiser2021_LiWD} because they have nearly identical volatility \citep{Lodders2003_abund}.
If a piece of continental crust were sufficiently heated to decrease log(K/Ca) by $\approx0.9$~dex to the level of \KCaSSP\ of WD~J1644–0449, then log(Na/Ca) would correspondingly decrease by $\approx0.9$~dex to a new $\log(\text{Na/Ca})\approx-0.9$. This new log(Na/Ca) would be just within 3$\sigma$ of the \NaCaSSP~of WD~J1644–0449. If the accretion is in decreasing phase, $\log(\text{Na/Ca})_{\text{DP}}$ of the planetesimal accreted by WD~J1644–0449 is easily compatible with a substantially thermally altered piece of continental crust.
However, the abundances of WD~J1644–0449, namely the log(K/Ca), strongly point away from the accreted planetesimal being comprised of unaltered continental crust material.

\subsection{SDSS~J1330+6435}\label{sec:J1330}
\subsubsection{SDSS~J1330+6435 Big Bang and Galactic Nucleosynthesis}\label{sec:J1330_bbn}

The planetesimal accreted by SDSS~J1330+6435 is compatible with formation in a protoplanetary nebula enhanced in log(Li/Ca) relative to the Solar System. Unfortunately, the limited abundance measurements and large error bars hardly constrain the circumstances as shown in Figure \ref{fig:allel_vs_Ca_ssp}. The \LiCaSSP\ of SDSS~J1330+6435 is within 3$\sigma$ of the [Fe/H]~=~–3.0, –2.0, and –1.0 bins of SAGA stars and CI Chondrites as is the \NaCaSSP. The \KCaSSP\ upper limit of SDSS~J1330+6435 is also compatible with CI Chondrites from the Solar System and therefore likely with the roughly solar log(K/Ca) implied by \citet{Kobayashi2020_originofelements}.
%The uncertainties on the abundances of SDSS~J1330+6435 are so large that even Solar System CI Chondrites are compatible with the abundances and limits for the accreted planetesimal assuming steady-state phase.
The old age of SDSS~J1330+6435 ($9.1^{+1.1}_{-0.5}$~Gyr) and slightly ambiguous kinematic membership (primarily resulting from the overlap of the thick and thin disk) point toward a sub-solar metallicity protoplanetary nebula.
However, the Li/Ca enhancement need not be large ($\sim0.7$~dex) if the star is thin disk.
Conversely, the consistency with the thin disk points toward a metallicity within $\approx0.7$~dex of solar, which is close enough that there would hardly be a Li-enhancement in the log(Li/Ca) of the protoplanetary nebula  \citep[e.g.][]{Grisoni2019_GalLievo}.

\subsubsection{SDSS~J1330+6435 Geologic Differentiation}\label{sec:J1330_geo}
SDSS~J1330+6435 is also compatible with the accretion of a piece of continental crust.
As stated, SDSS~J1330+6435 is compatible with thin disk kinematics and a near-solar metallicity nebular enrichment.
In Figure \ref{fig:allel_vs_Ca_ssp} the \LiCaSSP\ and \NaCaSSP\ of the body accreted by SDSS~J1330+6435 fall within 3$\sigma$ of continental crust, and the \KCaSSP\ upper limit is compatible with continental crust as well. 
However, if we rely on the estimated total age of the white dwarf ($9.1^{+1.1}_{-0.5}$~Gyr) instead of kinematic membership, the metallicity of the nebula would likely be sub-solar.
Regardless, the abundances of the accreted material in this star are consistent with a white dwarf that accreted continental crust.

\subsection{WD~J1824+1213}\label{sec:J1824}
\subsubsection{WD~J1824+1213 Big Bang and Galactic Nucleosynthesis}\label{sec:J1824_bbn}

WD~J1824+1213 has kinematics consistent with membership in the Galactic halo as demonstrated in Section \ref{sec:kinematics} and previously demonstrated \citep{Torres2019_WDKinematics, Hollands2021_LiWD, Elms2022_newLiWD, Bergeron2022_ultracoolWDsarenotcool}. This alone strongly indicates a sub-solar metallicity protoplanetary nebula from which the planetesimal accreted by WD~J1824+1213 formed. For example, in a sample of 36 kinematically-selected halo stars in the solar neighborhood, the median [Fe/H] was –1.3 with the middle 90\% of the stars falling in the range $-2.4<\text{[Fe/H]}<-0.6$ \citep{Bensby2014_generalabunds}.
Furthermore the halo is not believed to have undergone any Li production, so any departure from solar metallicity in the protoplanetary nebula of a halo object should appear proportionally in the starting Li/Ca, barring the $\sim0.3$ dex maximum deviation in $\alpha$-elements from solar values \citep{Bensby2014_generalabunds,Bensby2018_Liabunds}. Therefore, the starting Li/Ca abundance for the accreted planetesimal must have been higher than that of CI Chondrites in the Solar System, based on kinematic membership alone.

Therefore, at first, it would seem troubling that WD~J1824+1213 does not have the highest $\log(\text{Li/Ca})_{\text{SSP}}$ among the white dwarfs discussed in this work as it is part of the kinematic population with the lowest average metallicity as pointed out by \citet{Elms2022_newLiWD}.
But given the uncertainties in the ages and populations of the other stars and our lack of an age estimate for this one, the relatively lower Li/Ca is not ruled out.
%However, this assumes the white dwarfs are randomly drawn from their kinematic populations' respective metallicity distributions, which is not the case. All of the other white dwarfs with Li detections have total ages that place them among the oldest stars in the Galaxy (see Table \ref{tab:tot_ages} and Figure \ref{fig:tot_age_hist}), which suggests they should have a greater tendency toward sub-solar metallicity than the thin and thick disk on average. Furthermore as shown in Figure \ref{fig:allel_vs_Ca_ssp}, it is only WD~J2317+1830 that has a statistically significantly higher $\log(\text{Li/Ca})_{\text{SSP}}$ than that of WD~J1824+1214, so perhaps WD~J2317+1824 is simply one of the lowest metallicity thick disk stars while WD~J1824+1213 is one of the higher metallicity halo stars. Alternatively, WD~J2317+1830 may also be a halo star as shown in Figure \ref{fig:toomre}.
A more troubling issue is that the upper limit of \KCaSSP\ for WD~J1824+1213 falls below CI Chondrites as shown in Figure \ref{fig:allel_vs_Ca_ssp}.
Based on the implied constancy of log(K/Ca) over cosmological time from \citet{Kobayashi2020_originofelements}, we would need thermal alteration to account for the low log(K/Ca) in this star.
%Unfortunately, the SAGA Database \citep{Suda2017_SAGAIV} did not contain sufficient numbers of stars with simultaneous K, Ca, and Fe measurements to infer distributions of log(K/Ca) for low metallicity stars (or near-solar metallicity stars), so perhaps it is typical of halo stars to have sub-solar log(K/Ca). However, it is possible this is not the case, and log(K/Ca) remains roughly solar (and therefore CI Chondritic) through sub-solar metallicities.
%In this scenario, there must be an explanation for the lesser \KCaSSP~of WD~J1824+1213 that does not rely on nucleosynthesis alone. The potassium deficit in WD~J1824+1213 could be explained by some degree of heating of an otherwise primitive body.
CK Chondrites in the Solar System are depleted in log(Na/Ca) by 0.49~dex, log(K/Ca) by 0.55~dex, log(Li/Ca) by 0.29~dex, and log(Mg/Ca) by 0.09~dex relative to CI Chondrites \citep{Lodders1998_PSC}. The increasing phase abundances and limits of WD~J1824+1213 are consistent with CK Chondrites, save for log(Li/Ca), and the steady-state phase abundances for WD~J1824+1213 are all within 3$\sigma$ of the CK Chondrite abundances. 
Thus the abundances of the planetesimal accreted by WD~J1824+1213 are consistent with being a CK Chondrite analogue, with thermal alteration of a body formed from a primordial nebula with an enhanced log(Li/Ca) as expected for a star with halo kinematics.
%Nonetheless, based on the \KCaSSP~upper limit, the consistency of the other elemental abundances, and halo membership, we conclude that WD~J1824+1213 originated from a protoplanetary nebula with a Li excess. The body that polluted WD~J1824+1213 likely underwent heating resulting in a decreased abundance of volatiles, if the log(K/Ca) of halo stars is not sub-solar.

\subsubsection{WD~J1824+1213 Geologic Differentiation}\label{sec:J1824_geo}

As discussed in Section \ref{sec:J1824_bbn}, the lack of detectable K in the photosphere of WD~J1824+1213 sets an upper \KCaSSP\ limit that is too small to be compatible with continental crust. This problem cannot be resolved by diffusion timescales since they are so similar for each element. The only hypothesis that can be compounded with continental crust would be thermally processed continental crust that was tuned to remove Na and K and none of the less volatile lithium. Moreover, as a halo star, the effects of Big Bang and Galactic nucleosynthesis are inescapable in the primordial nebula from which WD~J1824+1213 formed.

\subsection{WD~J2317+1830}\label{sec:J2317}
\subsubsection{WD~J2317+1830 Big Bang and Galactic Nucleosynthesis}\label{sec:J2317_bbn}

WD~J2317+1830 has the highest \LiCaSSP\ of any of the white dwarfs under consideration as can be seen in Figure \ref{fig:allel_vs_Ca_ssp}. This extreme Li enhancement of $2.21\pm0.27$~dex compared to Solar System CI Chondrites is in line with the expected Galactic nucleosynthetic evolution of Li relative to Ca for the $\text{[Fe/H]}=-3.0$ bin ($-3.5<\text{[Fe/H]}<-2.5$), where Li is held at the BBN abundance as described in Section \ref{sec:saga_abunds}. The log(Na/Ca) of the main-sequence stars in this [Fe/H] bin ($\log(\text{Na/Ca})=-0.26\pm0.62$) with its large scatter is also compatible with the \NaCaSSP~of WD~J2317+1830 (\NaCaSSP~$=0.52\pm0.26$).
The high upper limit for \KCaSSP~for WD~J2317+1830 (\KCaSSP~$<0.92$) is also most likely compatible with the main-sequence log(K/Ca) abundance of this metallicity bin based on the [K/Fe] and [Ca/Fe] trends from \citet{Kobayashi2020_originofelements}, given that the log(K/Ca) of Solar System CI Chondrites is 2~dex less than this upper limit.
Our analysis excludes decreasing phase for WD~J2317+1830 following \citet{Hollands2021_LiWD} as its infrared excess indicative of a dust disk and relatively short diffusion timescales strongly suggest ongoing accretion. A sub-solar metallicity is also to be expected from the total age of WD~J2317+1830 ($7.7^{+0.3}_{-0.4}$~Gyr) and its likely thick disk (or possibly even halo) kinematic population membership, which further supports the Li enhancement originating from the combination of Big Bang and Galactic nucleosynthesis.

\subsubsection{WD~J2317+1830 Geologic Differentiation}\label{sec:J2317_geo}
WD~J2317+1830 exhibits some traits that are compatible with the accretion of continental crust, but overall it is not compatible with the accretion of ordinary continental crust material.
The \NaCaSSP~of WD~J2317+1830 (\NaCaSSP~$=0.52\pm0.26$) is compatible with continental crust, as is its high upper limit of \KCaSSP~(\KCaSSP~$<0.92$) as shown in Figure \ref{fig:allel_vs_Ca_ssp}. Its likely thick disk kinematic membership can be compatible with a solar abundance protoplanetary nebula; however, its advanced age ($7.7^{+0.3}_{-0.4}$~Gyr) likely puts it in the epoch of the thick disk for which an age-metallicity trend is present according to \citet{Bensby2014_generalabunds}. There is a declining trend of metallicity with age for stars older than $\approx8$~Gyr in the thick disk \citep{Bensby2014_generalabunds}. Therefore, in addition to the general skew of the thick disk toward lower metallicities, this advanced age also means the protoplanetary nebula of WD~J2317+1830 likely had a sub-solar metallicity. This makes it unlikely that a planet with abundances similar to Earth would form with correspondingly Earth-like continental crust. This is borne out by the 1.74~dex enhancement of the \LiCaSSP~of WD~J2317+1830 compared to continental crust as shown in Figure \ref{fig:allel_vs_Ca_ssp}. Thus, at minimum, if continental crust were accreted by WD~J2317+1830, it must have been a quite Li-rich piece of the crust. 

\citet{Hollands2021_LiWD} posited that the accreted material in WD~J2317+1830 was ``particularly lithium-rich crust.'' Areas of crust used for commercial lithium extraction represent the most Li-rich, accessible regions of the terrestrial crust.
Based on Li mines, we find that there are regions of the crust sufficiently enhanced in Li to meet the demands of this hypothesis from \citet{Hollands2021_LiWD}, although these Li-rich regions represent a very small fraction of the surface of the Earth's crust.
At $0.7\%$ lithium by mass \citep{Kesler2012_LiReserves}, a typical lithium deposit would have an adjusted log(Li/Ca) abundance from the continental crust \citep{Rumble2019_contcrust} of $\log\text{(Li/Ca)}_{\text{mine}}=-0.01$, which is a higher log(Li/Ca) abundance than the inferred \LiCaSSP~for the body accreted by WD~J2317+1830, $\log(\text{Li}/\text{Ca})_{\text{SSP}}=-0.82\pm0.27$.
%Therefore there is some rock in the Earth's continental crust that is more Li-rich than the material accreted by WD~J2317+1830.
Adapting equation 4 from \citet{Hollands2018_DZII} to produce equation \ref{eq:li_dilution}, we calculated that $\approx15\%$ of the accreted continental crust would have to be lithium mining deposits: 
\begin{equation}\label{eq:li_dilution}
    \text{M}_{\text{mine}}=\frac{(N_{\text{Li}}/N_{\text{Ca}})_{\text{SSP}}(\mu_{\text{Li}}/\mu_{\text{Ca}})\text{Cru}_{\text{Ca}}-\text{Cru}_{\text{Li}}}{(\text{Mine}_{\text{Li}}-\text{Cru}_{\text{Li}})-(N_{\text{Li}}/N_{\text{Ca}})_{\text{SSP}}(\mu_{\text{Li}}/\mu_{\text{Ca}})(\text{Mine}_{\text{Ca}}-\text{Cru}_{\text{Ca}})},
%\text{M}_{\text{mine,tot}}=\frac{(N_{\text{Li}}/N_{\text{Ca}})_{\text{SSP}}(\mu_{\text{Li}}/\mu_{\text{Ca}})\text{M}_{\text{crust,Ca}}-\text{M}_{\text{crust,Li}}}{(\text{M}_{\text{mine,Li}}-\text{M}_{\text{crust,Li}})-(N_{\text{Li}}/N_{\text{Ca}})_{\text{SSP}}(\mu_{\text{Ca}}/\mu_{\text{Li}})(\text{M}_{\text{mine,Ca}}-\text{M}_{\text{crust,Ca}})}.
\end{equation} %see page 65-66 of postdoc II for the by-hand derivation
where $\text{M}_{\text{mine}}$ is the fraction of accreted material comprised of lithium mining deposits, $(N_{\text{Li}}/N_{\text{Ca}})_{\text{SSP}}$ is the relative number abundance of those elements in the accreted material assuming steady-state phase (same as 10 to the power of \LiCaSSP), $\mu_{\text{el}}$ is the atomic mass of an element, and $\text{Cru}_{\text{el}}$ and $\text{Mine}_{\text{el}}$ are the mass fractions of continental crust and lithium mining deposits that are comprised of a given element.
In contrast to the $\approx$~15\% requirement, only $\approx10^{-6}$ of the total Li in the continental crust of Earth is found in these Li deposits \citep{Rudnick2003_continentalcrust,Rumble2019_contcrust}.
%Continuing the lithium mine accretion scenario, we calculated the total mass of Li in the convection zone of WD~J2317+1830 to be 290 megatons using the log(Li/He) and convection zone mass ($\text{M}_{\text{CVZ}}$) of \citet{Hollands2021_LiWD}. By contrast, the Earth's total available global lithium resources (i.e. the total mass of minable lithium) are estimated to only total 31.1 megatons \citep{Kesler2012_LiReserves}. It would seem WD~J2317+1830 would have to have already accreted all of the minable lithium of Earth's continental crust nearly ten times over to hold its present lithium mass. However, the total lithium mass in the continental crust of the Earth can be estimated to be $8\times10^8$ megatons  assuming a total continental crust mass of $4\times10^{25}$ g \citep{Rudnick2003_continentalcrust}, and a lithium concentration of 20 ppm \citep{Rumble2019_contcrust}, so if there are more lithium-rich sections of the crust deeper in the crust than mining can reach, perhaps it could be done.
%Would this Li-rich material be ``continental crust?'' Nearly all of the continental crust is not this enriched in Li. Therefore, if WD~J2317+1830 were accreting material that is substantially lithium mining material, would it not then be something other than continental crust?
Thus the enhancement in lithium is not compatible with terrestrial continental crust, though we cannot rule out an exotic planet that might have larger or more enriched lithium deposits.

\subsection{LHS~2534}\label{sec:LHS2534}
\subsubsection{LHS~2534 Big Bang and Galactic Nucleosynthesis}\label{sec:LHS2534_bbn}

LHS~2534 is not compatible with the simple Big Bang lithium enhancement in an old system.
%LHS~2534 is most likely a member of the thick disk or potentially the halo (see Section \ref{sec:kinematics}), both of which suggest a sub-solar metallicity as would be required for the combination of Big Bang and Galactic Nucleosynthesis to cause the \LiCaSSP\ excess of the accreted body. The age of LHS~2534 ($8.1^{+2.3}_{-0.8}$~Gyr) is also compatible with thick disk membership and sub-solar metallicity.
The kinematics (thick disk or halo) and age ($8.1^{+2.3}_{-0.8}$~Gyr) of LHS~2534 suggest a sub-solar metallicity.
The \LiCaSSP\ is compatible with the $\text{[Fe/H]}=-2.0$ and $-3.0$ bins. The large scatter of log(Na/Ca) in SAGA stars in the $\text{[Fe/H]}=-2.0$ and $-3.0$ bins overlap the \NaCaSSP\ of LHS~2534. The \FeCaSSP\ of LHS~2534 is also compatible with the $\text{[Fe/H]}=-2.0$ and $-3.0$ bins. 

However, \KCaSSP\ and \CrCaSSP\ are incompatible with the $\text{[Fe/H]}=-2.0$ and $-3.0$ bins \citep[implied by][in the case of log(K/Ca)]{Kobayashi2020_originofelements}, and the accretion phase would not correct either of these discrepancies as shown in Figure \ref{fig:allel_vs_Ca_ssp}. K and Ca diffuse at essentially the same rate, so the inferred log(K/Ca) is the same regardless of phase. Cr diffuses more slowly than Ca, which means decreasing phase increases the inferred log(Cr/Ca) of the accreted body, moving it further from the log(Cr/Ca) of the $\text{[Fe/H]}=-2.0$ and $-3.0$ bins. Thermal alteration would also decrease log(K/Ca) and log(Cr/Ca) rather than make them overabundant because Ca would be the last element among the three to be depleted \citep{Lodders2003_abund}. The \KCaSSP\ and \CrCaSSP\ enhancements suggest geological or other processes at work.
Thus while the Li enhancement is what we expect from Big Bang and Galactic nucleosynthesis, the K and Cr are unexplained in this model.

\subsubsection{LHS 2534 Geologic Differentiation}\label{sec:LHS2534_geo}
LHS~2534 is not compatible with the accretion of ordinary continental crust material. As shown in Figure \ref{fig:allel_vs_Ca_ssp}, the \LiCaSSP~of LHS~2534 is within 3$\sigma$ of continental crust, so the Li excess is within the range of geological effects. The \NaCaSSP~and \KCaSSP~of LHS~2534 are also within 3$\sigma$ of continental crust, so the alkali metals appear to be present in ratios consistent with continental crust. Additionally, due to the similarity of the diffusion timescales of K and Ca, this consistency with continental crust should hold regardless of the accretion phase.

However, LHS~2534's membership in the thick disk (or potentially halo, see Figure \ref{fig:toomre}) and advanced age ($8.1^{+2.3}_{-0.8}$~Gyr), both point away from a solar metallicity for the protoplanetary nebula as would be expected in order to produce an Earth-like planet and subsequently Earth-like continental crust. Additionally, the steady-state phase abundances of the iron group elements, \CrCaSSP~and \FeCaSSP, are both more than 3$\sigma$ more abundant for LHS~2534 than in continental crust as shown in Figure \ref{fig:allel_vs_Ca_ssp}. Both of these inferred abundances increase the further into decreasing phase accretion is assumed to be as well, so the accretion phase uncertainty is not able to improve compatibility.  The \CrCaSSP\ is particularly problematic as there is more than 2~dex too much Cr relative to Ca for the accreted material to be continental crust. Therefore, the measurements of LHS~2534 do not support the accretion of continental crust.

\subsection{SDSS~J1636+1619}\label{sec:SDSSJ1636}

%We suspect the null detections of Li, Na, and K in SDSS~J1636+1619 are the result of the accreted body having undergone substantial heating causing the far more volatile alkalis to be largely removed from the planetesimal at some point before accretion by the white dwarf. The resulting planetesimal retained only the more refractory elements, such as Ca, which was then accreted by the WD resulting in a photosphere in which only Ca is present at detectable levels. Unfortunately, this makes it rather useless in discerning which hypothesis could be correct.
\subsubsection{SDSS J1636+1619 Big Bang and Galactic Nucleosynthesis}\label{sec:J1636_bbn}
SDSS~J1636+1619 most likely did not accrete a pristine, primitive planetesimal from a protoplanetary nebula at any metallicity.
The advanced age of SDSS~J1636+1619 ($9.3^{+2.2}_{-0.9}$~Gyr) suggests a sub-solar metallicity for the protoplanetary nebula that gave rise to the accreted planetesimal. Unfortunately, Ca was the only element for which a specific abundance could be measured in the photosphere of SDSS~J1636+1619, so we are forced to consider only upper limits in our abundance analysis. The upper limits of \LiCaSSP, \KCaSSP, and \FeCaSSP\ allow for the accretion of a primitive planetesimal from the $\text{[Fe/H]}=$~–2.0, –1.0, or 0.0 bin. However, the \NaCaSSP~upper limit restricts the accretion of a primitive planetesimal to the $\text{[Fe/H]}=$~–2.0 or –1.0 bins. The \MgCaSSP~upper limit further restricts the compatible primitive planetesimals to those formed in the $\text{[Fe/H]}=$~–2.0 bin. The likely thin disk kinematic membership of SDSS~J1636+1619, however, essentially eliminates the $\text{[Fe/H]}=$~–2.0 bin.
Alternatively, a planetesimal from any of the metallicity bins subjected to sufficient heating could be compatible with the abundance upper limits of SDSS~J1636+1619. Perhaps SDSS~J1636+1619 accreted a planetesimal from the $\text{[Fe/H]}=$~–1.0 or 0.0 bin that was subjected to substantial heating causing depletion of the elements more volatile than Ca (all of the other elements considered).

\subsubsection{SDSS J1636+1619 Geologic Differentiation}\label{sec:J1636_geo}
SDSS~J1636+1619 could not have accreted unaltered continental crust material.
The thin disk kinematics of SDSS~J1636+1619 suggest it may have had a solar metallicity protoplanetary nebula, which would have enabled the formation of an Earth-like planet and continental crust. The \LiCaSSP, \MgCaSSP, and \FeCaSSP~upper limits are compatible with the accretion of continental crust material. The \KCaSSP~upper limit is technically below the log(K/Ca) of the continental crust, but with the diffusion timescale uncertainties, it is reasonable to allow for a $\approx0.1$~dex mismatch between the central value upper limit and continental crust. Therefore, we consider the \KCaSSP~upper limit compatible with continental crust for SDSS~J1636+1619. However, the \NaCaSSP~upper limit is $\approx0.6$~dex less than that of continental crust, so that is incompatible with the accretion of continental crust in this case. As with the Big Bang and Galactic nucleosynthetic explanation in Section \ref{sec:J1636_bbn}, SDSS~J1636+1619 could have accreted continental crust that was subjected to substantial heating such that it depleted the elements more volatile than Ca (all of the other elements discussed), but it could not have accreted pristine continental crust like that of Earth.

\section{Caveats and Lurking Issues}\label{sec:caveats}

\subsection{Thick versus Thin Hydrogen Envelopes and WD~J2317+1830}\label{sec:H-layer}
%The cooling ages of white dwarfs are heavily dependent upon envelope composition \citep[e.g.][]{Bedard2020_specevoI}. In Section \ref{sec:WD_ages}, we derived cooling ages for the white dwarfs by adopting probabilities for envelope compositions instead of assuming a given white dwarf has a ``thick'' or ``thin'' H mass as is usually done \citep[e.g.][]{Kaiser2021_LiWD,Hollands2021_LiWD}. Here we explain why we departed from the usual convention of choosing one envelope or the other.
The cooling ages of white dwarfs are heavily dependent upon the hydrogen mass ($\text{M}_{\text{H}}$) in their envelopes \citep[e.g.][]{Bedard2020_specevoI}. In Section \ref{sec:WD_ages}, we derived cooling ages for all of the white dwarfs using ``thin'' H mass ($\text{M}_{\text{H}}=10^{-10} \text{M}_\odot$). Here we explain why we opted to use the ``thin'' H mass for all white dwarfs instead of using the ``thick'' H mass ($\text{M}_{\text{H}}=10^{-4} \text{M}_\odot$) for WD~J2317+1830 as was done by \citet{Hollands2021_LiWD}.
Our total age for WD~J2317+1830 ($7.7^{+0.3}_{-0.4}$~Gyr) using the ``thin'' H mass is $\approx 2$~Gyr less than the total age \citet{Hollands2021_LiWD} found using the ``thick'' H mass ($9.7\pm0.2$~Gyr).
%We demonstrate how to calculate a hydrogen mass for individual white dwarfs leading us to conclude WD~J2317+1830 is most appropriately described using a ``thin'' H mass (as are the other white dwarfs).
Using WD~J2317+1830 as an exemplar, we present below the method we use to determine the total H in the envelope and show that it is more appropriately described by the thin H cooling models, a result that applies to all stars in our sample.

We can use the convection zone masses ($M_{\text{CVZ}}$) and the H/He abundances to calculate the total hydrogen mass in the convection zone of the white dwarfs and treat that as the total hydrogen mass. The equation to calculate $\text{M}_{\text{H}}$ this way is the following:
\begin{equation}\label{eq:MH_nohidden}
\text{M}_{\text{H}}=\frac{\text{M}_{\text{CVZ}}/\text{M}_{\text{WD}}}{1+\Big(\frac{\mu_{\text{He}}}{\mu_{\text{H}}}\Big)10^{-\log(\text{H}/\text{He})}}\times\text{M}_{\text{WD}}
,
\end{equation}
where $\mu_{\text{el}}$ is the atomic mass of an element, and  we assume H and He are the dominant elements by mass in the convection zone.

Directly computing the hydrogen mass in the convection zone for WD~J2317+1830 with its near-equal abundances of H and He (from atmospheric model fits, see Table \ref{tab:atm_params}) using Equation \ref{eq:MH_nohidden}, yields $\text{M}_{\text{H}}=2\times10^{-9} \text{M}_\odot$ of hydrogen, placing it much closer to the $10^{-10} \text{M}_\odot$ hydrogen mass of the ``thin'' hydrogen envelope cooling models. This is in contrast with the value we obtain for WD~J1824+1213, $\text{M}_{\text{H}}=2\times10^{-6} \text{M}_\odot$,
despite the nearly identical H/He photospheric abundances of the two white dwarfs. This discrepancy is borne from the disparate convection zone masses of WD~J2317+1830 and WD~J1824+1213, $\log(M_{\text{CVZ}}/M_{\text{WD}})=-7.7$ and $\log(M_{\text{CVZ}}/M_{\text{WD}})=-4.2$, respectively.
%However, if we consider a white dwarf with a thick hydrogen envelope, $\text{T}_{\text{eff}}$ = 5000 K, and log(g)=8.5 (the highest mass and lowest $\text{T}_{\text{eff}}$ available in \citep{Koester2020_WDenvelope}), the convection zone mass is $\sim10^{-7} \text{M}_\odot$. As this convection zone is solely hydrogen, $\text{M}_{\text{H}}=\text{M}_{\text{CVZ}}$, a white dwarf that has the maximum amount of hydrogen in its convection zone appears to have only $\sim10^{-7} \text{M}_\odot$ of hydrogen using Equation \ref{eq:MH_nohidden}. Thus, the hydrogen mass contained in the convection zone of a white dwarf is at best only a lower limit on the total hydrogen mass, so the correct expression for the hydrogen mass in the case of a pure hydrogen atmosphere would be $\text{M}_{\text{H}} \geq \text{M}_{\text{CVZ}}$.
We present the hydrogen masses or upper limits ($\text{M}_{\text{H}}$) as calculated using Equation \ref{eq:MH_nohidden} for the white dwarfs discussed in this work in Table \ref{tab:MH}.
\begin{deluxetable*}{lcR}[!ht]
\tablecaption{Inferred hydrogen masses for the white dwarfs discussed in this work based on Equation \ref{eq:MH_nohidden} and convection zone masses ($M_{\text{CVZ}}$).\label{tab:MH}}
\tablewidth{0pt}
\tablehead{\colhead{Object} & \colhead{$\log(M_{\text{CVZ}}/M_{\text{WD}})$} & \colhead{$\text{M}_{\text{H}}\text{ (M}_{\odot}\text{)}$}
}
    \startdata
        WD~J1644$-$0449 & -4.4\tablenotemark{a} & <5\times10^{-8} \\ 
        SDSS~J1330+6435 & N/A & \text{N/A} \\ 
        WD~J1824+1213 & -4.2 & 2\times10^{-6} \\
        WD~J2317+1830 & -7.7 & 2\times10^{-9}  \\ 
        LHS~2534 & -5.5& <4\times10^{-10} \\ 
        %WD~J2356$-$209 & <-2.3 & 1.1\pm0.2 & 1.4\pm0.22& <-0.95 & <-0.5 & 0.8\pm0.22 \\
        SDSS~J1636+1619 &-5.4& <7\times10^{-10}\\ 
    \enddata
\tablenotetext{a}{\citet{Kaiser2021_LiWD}}
\end{deluxetable*}

\citet{Rolland2018_specevo} explored the convection zone masses of thin H-atmosphere white dwarfs. We can alter our existing Equation \ref{eq:MH_nohidden} following Equation 4 of \citet{Rolland2018_specevo} to include the amount of hydrogen concealed below the convection zone, $R$. The new form of the equation is given by the following:
\begin{equation}\label{eq:MH_somehidden}
\text{M}_{\text{H}}=\frac{\text{M}_{\text{CVZ}}/\text{M}_{\text{WD}}}{1+\Big(\frac{\mu_{\text{He}}}{\mu_{\text{H}}}\Big)10^{-\log(\text{H}/\text{He})}}\times\text{M}_{\text{WD}}(1+R).
\end{equation}
\citet{Rolland2018_specevo} suggested the simple use of $R=2$, indicating two-thirds of the hydrogen is below the convection zone. These white dwarfs would need orders of magnitude more hydrogen to be concealed below the convection zone to reach the ``thick'' hydrogen mass, so this seems unlikely to change which cooling models are more appropriate. However, \citet{Rolland2018_specevo} only probed down to 6000 K and treated all white dwarfs as having a mass of 0.6 $\text{M}_{\odot}$, so the expected atmospheric compositions for these hydrogen masses at these low temperatures and different masses are not completely theoretically explored.

Future efforts in estimating total ages would benefit from cooling model grids that include multiple hydrogen masses and more importantly a means of determining total hydrogen mass based on surface compositions. This could be achieved via predictions of the chemical structure from the white dwarf evolution models. \citet{Bedard2022_specevoII} modeled the chemical structure of a 0.6~$\text{M}_\odot$ white dwarf with  $\text{M}_{\text{H}}=10^{-10} \text{M}_\odot$ from $T_{\text{eff}}$=90,000~K down to $T_{\text{eff}}$=8,183~K. The models will need to be extended to lower temperatures, additional white dwarf masses, and to include at least both canonical hydrogen masses ($10^{-4} \text{M}_\odot$ and $10^{-10} \text{M}_\odot$), but a hydrogen layer mass grid would probably be more beneficial.

%\textbf{***What did Koester 2020 assume as the content of the atmosphere for his convection zones in his paper? He must have assumed some relative H/He abundances because Simon described the convection zone calculation in an email on 2021-11-19 (subject: "Research meeting w/ Galactic lithium evolution modeler"). Simon indicated Rolland (and Simon himself) assume the convection zone is the abundance of the photosphere and then calculate how deep convection would develop in that environment. So Koester would have to either calculate (or assume) a H/He abundance in each scenario for the convection zone mass determination. And probably also to get the diffusion timescales as it affects the ionization due to pressure and stuff I think?}

\subsection{Magnetism and LHS 2534}\label{sec:magnetism}

%Numerous authors have suggested that magnetism can suppress convection in white dwarfs \citep{Wickramasinghe1986_magnetism, Valyavin2014_magnetism, Tremblay2015_convection}, and \citet{Tremblay2015_convection} determined the suppression should occur in white dwarfs with magnetic field strengths $\gtrsim$~50~kG.
%\citet{GentileFusillo2018_magnetic_convection} confirmed that atmospheric convection was suppressed in a hydrogen-dominated, magnetic white dwarf, WD 2105-820, which \citet{Landstreet2012_magneticDA} determined to have a polar magnetic field strength of $\simeq$~56~kG. LHS~2534 has a magnetic field strength of 2.12 $\pm$ 0.03 MG \citep{Hollands2017_DZI}, which is $\sim$40 times the predicted minimum field strength to suppress convection \citep{Tremblay2015_convection}, so it is highly likely that convection in LHS~2534 is suppressed. The diffusion timescales used to compute the abundances of the accreted planetesimals for steady-state and decreasing phases in Section \ref{sec:infer_abund} are computed for the base of the convection zone of a white dwarf with the given atmospheric parameters because the metals are efficiently mixed inside the convection zone. \citet{Hollands2021_LiWD} calculated that the convective velocities would be damped by less than one order of magnitude. If LHS~2534 has lesser convection, its diffusion timescales should instead be calculated for the base of this revised convection zone (or even the base of the photosphere, which our diffusion timescale calculations did not account for.

The modeling of LHS~2534 was particularly plagued by peculiarities that are most likely related to magnetism given that is the defining feature of LHS~2534. There are several likely Ca I absorption features near 5590, 6120, and 6450 Å that are not modeled with the present abundances. However, to match the strengths of these lines, the Ca abundance or \Teff~of the white dwarf needs to be adjusted upward enough that the broadband SED no longer is fit by the model. This fundamental mismatch of parameters may result from the lack of a unified treatment of magnetism and atmospheric pressure effects. Alternatively the presence of star spots may lead to this issue. LHS~2534 exhibits photometric variability consistent with a star spot (Hermes private communication). Spots on white dwarfs at these lower \Teff\ values are generally thought to be a lower temperature than the surrounding surface due to the suppression of convection \citep[e.g.][]{Reding2018_K2spottedWDs}.
\citet{Bagnulo2024_variablemetalsDZH} and \citet{Bagnulo2024_accretionscars_metalvar} recently identified time-variable metal abundances in two other low-\Teff\ DZHs, WD~0816–310 and WD~2138$-$332, which they attributed to spatially-variable metal abundances on their surfaces. \citet{Bagnulo2024_variablemetalsDZH} and \citet{Bagnulo2024_accretionscars_metalvar} suspect the inhomogeneous surface distribution of metals was caused by magnetically funneled accretion onto the magnetic poles and poor horizontal mixing due to the suppression of convection by magnetism. While WD~0816–310 and WD~2138$-$332 are hotter (6250~K and 7150~K) than LHS~2534 ($5020\pm100$~K), their maximum magnetic field strengths \citep[$\approx 48$~kG and $\approx 14$~kG,][]{Bagnulo2024_variablemetalsDZH,Bagnulo2024_accretionscars_metalvar} are much less than the magnetic field of LHS~2534 \citep[$2.12\pm0.03$~MG from][]{Hollands2017_DZI}, so it is reasonable to expect a similar phenomenon to occur. 
Our models assume a homogeneous surface in both abundances and \Teff, so spots are not handled by the models.
Therefore these parameters are all suspect for LHS~2534. This would be the case for the models used by \citet{Hollands2021_LiWD} as well.

%\textbf{\citep[($\approx 48$~kG and $\approx 14$~kG)][]{Bagnulo2024_variablemetalsDZH,Bagnulo2024_accretionscars_metalvar} are}

\subsection{Accretion Phases and Diffusion Timescales}\label{sec:caveats_accretion_diffusion}

In addition to potential issues with magnetic fields impacting the convection zone depth and therefore the conditions in which the diffusion timescales should be computed, the diffusion timescales are calculated in a less than ideal way. Diffusion coefficients are used to infer the abundances of the accreted bodies for steady-state and decreasing phase as shown in Sections \ref{sec:steadystate} and \ref{sec:decreasing}. Therefore the planetesimal abundances inferred for each white dwarf we analyzed would be different if our diffusion timescales are inaccurate.

In addition to being calculated via extrapolation of a model grid in our case, even the bespoke diffusion timescales like those used in \citet{Hollands2021_LiWD} rely on the diffusion coefficients of \citet{Paquette1986_diffusioncoefficients}. \citet{Heinonen2020_betterdiffusion} recently presented a superior method of computing these diffusion coefficients, which revealed a factor of $\gtrsim3$ difference with the relative diffusion timescales of Ca and Si calculated by \citet{Paquette1986_diffusioncoefficients}. The diffusion coefficients for the other elements have not been calculated using the updated method of \citet{Heinonen2020_betterdiffusion}, but it is likely the diffusion coefficients of the other elements will differ as well.

\subsection{Three More Low-Temperature, metal-polluted white dwarfs (Two with Li Pollution)}\label{sec:elms_objects}

Late in the drafting of this paper, \citet{Elms2022_newLiWD} published their findings for two more low-\Teff~white dwarfs with metal pollution: WD~J2147–4035 (Li) and WD~J1922+0233 (no Li). Subsequently, \citet{Vennes2024_magLiWD} found another white dwarf with Li pollution, 2MASS~J0916$-$4215. Both \citet{Elms2022_newLiWD} and \citet{Vennes2024_magLiWD} concluded that their Li-polluted white dwarfs were consistent with continental crust.
Notably, both WD~J2147–4035 and 2MASS~J0916$-$4215 were found to be magnetic like LHS~2534, and WD~J2147–4035 was found to exhibit photometric variability indicative of spots also like LHS~2534 \citep{Elms2022_newLiWD,Vennes2024_magLiWD}. 2MASS~J0916$-$4215 may also have a spot aligned with its spin axis, which would prevent photometric variability while still causing an inhomogeneous photospheric temperature that would confound modeling.
These objects were excluded from our analysis because we wanted all of the abundances to be based on the same model to avoid model-dependent issues, and spots and magnetism are not adequately modeled presently in any case.

\section{Conclusions}\label{sec:conclusions}

We presented new spectroscopic observations of three Li-polluted white dwarfs: WD~J1824+1213, WD~J2317+1830, and LHS~2534. We also presented new spectroscopic observations of the low-\Teff~DZ SDSS~J1636+1619. We provided atmospheric model parameters for these white dwarfs using state-of-the-art white dwarf models for low \Teff.

In the context of these white dwarfs and two additional Li-polluted white dwarfs (WD~J1644–0449 and SDSS~J1330+6435) previously analyzed with the same atmospheric models, we evaluated three hypotheses in the literature put forward to explain the excess of Li observed in white dwarfs compared to CI Chondrites in the Solar System:
\begin{enumerate}
\item{Big Bang and Galactic Nucleosynthesis \citep{Kaiser2021_LiWD}}
\item{Geologic Differentiation \citep{Hollands2021_LiWD}}
\item{Icy Exomoons Formed from the Spalled Icy Rings of Gas Giant Planets \citep{Doyle2021_spallWD}.}
\end{enumerate}

We determined the exomoon spallation hypothesis of \citet{Doyle2021_spallWD} requires a host star to survive at least 10 times the lifetime of WD~J2317+1830's progenitor with appropriate reaction cross sections and proton flux of Jupiter. Furthermore, \citet{Trierweiler2022_exomoonfrequency} determined exomoons are highly unlikely to be accreted. Therefore, we conclude that the accretion of exomoons comprised of spalled icy ring material is unlikely to be the source of the Li excess in white dwarfs.

We determined that WD~J644–0449 is compatible with the accretion of a primitive planetesimal from the $-2.5\leq\text{[Fe/H]}\leq-0.5$ bins or a Solar System CI Chondrite, which supports a combination of Big Bang and Galactic nucleosynthetic origin of the Li excess.
SDSS~J1330+6435 is compatible with the accretion of a primitive planetesimal formed from protoplanetary nebulae with $-3.5<\text{[Fe/H]}<-0.5$ or a normal CI Chondrite.
WD~J2317+1830 can be explained by the accretion of a primitive planetesimal formed from a protoplanetary nebula with $-3.5<\text{[Fe/H]}<-2.5$, in line with this hypothesis. WD~J1824+1213 and SDSS~J1636+1619 could be explained by the accretion of planetesimals that were subjected to devolatilization via heating, but in the case of WD~J1824+1213, its halo kinematics indicate the protoplanetary nebular metallicity would most likely have been sub-solar and therefore impacted by Big Bang and Galactic nucleosynthesis. LHS~2534 as modeled, however, cannot be squared with a Big Bang and Galactic nucleosynthetic origin of its Li excess without some other process (other than heating) at play.

We found that only SDSS~J1330+6435 is compatible with the accretion of unaltered continental crust, but with its large errorbars, it is compatible with many Solar System bodies. WD~J1644–0449 and SDSS~J1636+1619 could be compatible with the accretion of heavily thermally altered continental crust. WD~J1824+1213 could be compatible with heavily thermally altered continental crust, were it not for its halo kinematics and implicit sub-solar metallicity. WD~J2317+1830 and LHS~2534 cannot be squared with the accretion of ordinary continental crust even if it were subjected to heating.

%The evaluation of WD~J1824+1213 and WD~J2317+1830 are most salient for determining which of these hypotheses is likely correct. They each could only be explained when considering the effects on relative abundances in the protoplanetary nebulae resulting from Big Bang and Galactic nucleosynthesis. LHS~2534 defies any explanation presented thus far. 
Perhaps LHS~2534 has recently accreted from two different bodies (Hollands private communication) as proposed by \citet{Johnson2022_doubleaccretion_nitrogen} for another white dwarf. It is more likely the case that its magnetism is in some way sabotaging our efforts to model its atmospheric parameters.

Future work would benefit from the unified treatment of magnetism and high pressure effects in cool white dwarf atmospheres. Total age calculations would benefit from a cooling model grid that includes H-masses between the canonically ``thick'' and ``thin'' values. Computation of the diffusion coefficients and subsequent diffusion timescales with the method of \citet{Heinonen2020_betterdiffusion} down to the \Teff~$\approx3,000$~K would likely improve our abundance analysis as well. More measurements of K in extremely low metallicity main-sequence stars would also be of benefit in determining the Galactic evolution of K and therefore the expected abundances for ancient planetesimals accreted by white dwarfs. JHK IR photometry of WD~J1644–0449 and SDSS~J1330+6435 could enable a robust H/He determination, which would allow the MgH feature to be exploited to obtain log(Mg/Ca) and introduce another constraint on the accreted planetesimals. Higher signal-to-noise spectroscopic follow-up of SDSS~J1330+6435 with robust telluric corrections could also enable a stronger log(K/Ca) limit or a detection to better discriminate between continental crust and a combination of Galactic and Big Bang nucleosynthesis as the mechanism of its potential Li enhancement. There is also still the mystery of the overly narrow Li lines compared to the model atmospheres.

We find Big Bang and Galactic nucleosynthesis to be the most plausible explanation of the abundances in WD~J1644–0449, WD~J1824+1213, and WD~J2317+1830. SDSS~J1330+6435 will require stricter abundances to determine its planetesimal's origins, and LHS~2534, as presently modeled, defies all three hypotheses.

\begin{acknowledgements}
The authors would like to thank Mark Hollands for providing the spectra of WD~J1824+1213, WD~J2317+1830, and LHS~2534 from his work.
B.C.K. would like to thank J. Whitworth, D. Carr, and C. Poovey for helpful conversations and insights. B.C.K. would like to thank Steven Singletary and Nikos Prantzos for helpful conversations about lithium in geology and Galactic nucleosynthesis. B.C.K. would like to thank Mark Hollands for helpful conversations about these white dwarfs.

B.C.K. was supported by the NC Space Grant Graduate Research Fellowship.
S.B. was supported by a Banting Postdoctoral Fellowship and a CITA (Canadian Institute for Theoretical Astrophysics) National Fellowship. This work was supported by funding from the National Science Foundation under grant number AST-2108311.

Based on observations obtained at the Southern Astrophysical Research (SOAR) telescope, which is a joint project of the Minist\'{e}rio da Ci\^{e}ncia, Tecnologia e Inova\c{c}\~{o}es (MCTI/LNA) do Brasil, the US National Science Foundation’s NOIRLab, the University of North Carolina at Chapel Hill (UNC), and Michigan State University (MSU).
This work has made use of data from the European Space Agency (ESA) mission {\it Gaia} (\url{https://www.cosmos.esa.int/gaia}), processed by the {\it Gaia} Data Processing and Analysis Consortium (DPAC, \url{https://www.cosmos.esa.int/web/gaia/dpac/consortium}). Funding for the DPAC has been provided by national institutions, in particular the institutions participating in the {\it Gaia} Multilateral Agreement.
Based on observations obtained at the international Gemini Observatory, a program of NSF NOIRLab, which is managed by the Association of Universities for Research in Astronomy (AURA) under a cooperative agreement with the U.S. National Science Foundation on behalf of the Gemini Observatory partnership: the U.S. National Science Foundation (United States), National Research Council (Canada), Agencia Nacional de Investigaci\'{o}n y Desarrollo (Chile), Ministerio de Ciencia, Tecnolog\'{i}a e Innovaci\'{o}n (Argentina), Minist\'{e}rio da Ci\^{e}ncia, Tecnologia, Inova\c{c}\~{o}es e Comunica\c{c}\~{o}es (Brazil), and Korea Astronomy and Space Science Institute (Republic of Korea).

\end{acknowledgements}

\software{astropy \citep{astropy2013, astropy2018, astropy2022}}

\appendix

\section{Model Fit Plots for White Dwarfs}\label{sec:model_fit_plots}

\begin{figure*}[htb!]
\includegraphics[scale=1]{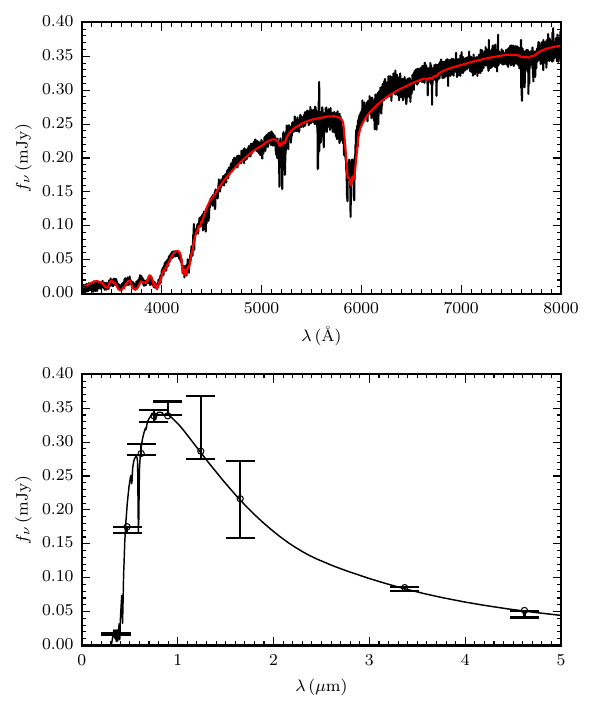}
\centering
\caption{LHS~2534 model fit from Section \ref{sec:atm_modeling} to VLT/X-Shooter Data from \citet{Hollands2021_LiWD} and photometric points. \label{fig:LHS2534_xshooter_model}}
\end{figure*}

\begin{figure*}[htb!]
\includegraphics[scale=1]{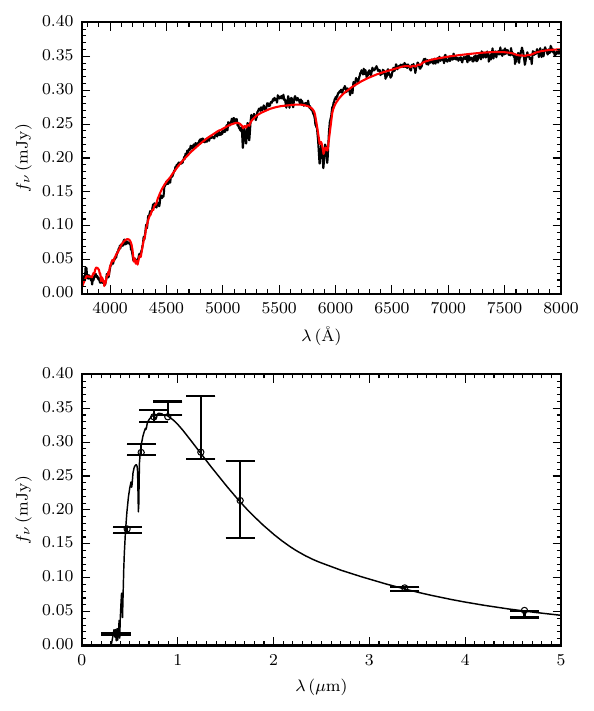}
\centering
\caption{LHS~2534 model fit from Section \ref{sec:atm_modeling} to SOAR/Goodman Spectrograph Data from Section \ref{sec:obs} and photometric points. \label{fig:LHS2534_goodman_model}}
\end{figure*}

\begin{figure*}[htb!]
\includegraphics[scale=1]{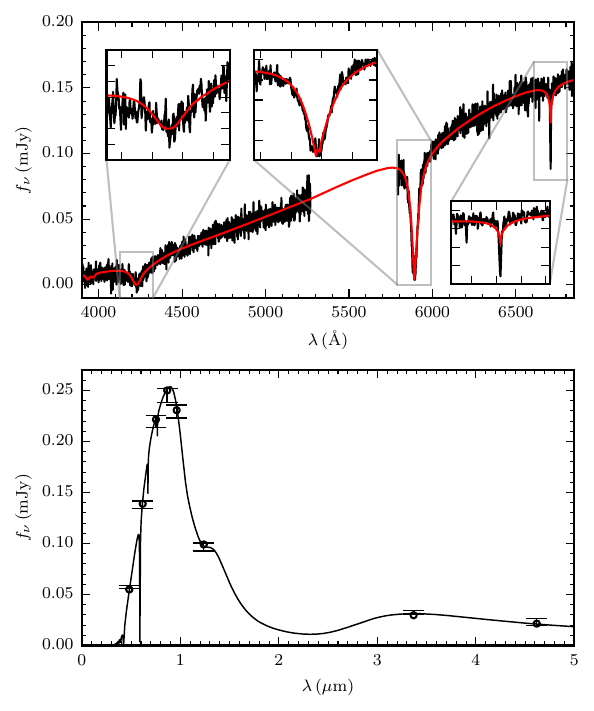}
\centering
\caption{WD~J1824+1213 model fit from Section \ref{sec:atm_modeling} to WHT/ISIS Data from \citet{Hollands2021_LiWD} and photometric points. \label{fig:J1824_isis_model}}
\end{figure*}

\begin{figure*}[htb!]
\includegraphics[scale=1]{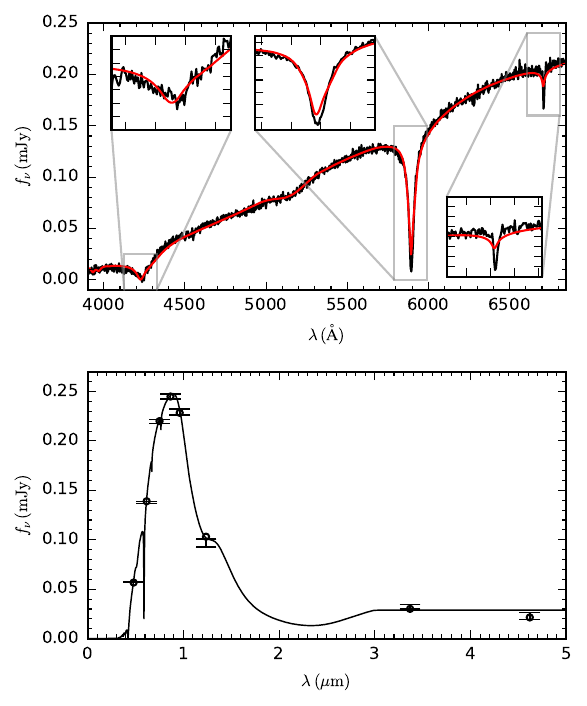}
\centering
\caption{WD~J1824+1213 model fit from Section \ref{sec:atm_modeling} to Soar/Goodman Spectrograph Data from Section \ref{sec:obs} and photometric points. \label{fig:J1824_goodman_model}}
\end{figure*}

\begin{figure*}[htb!]
\includegraphics[scale=1]{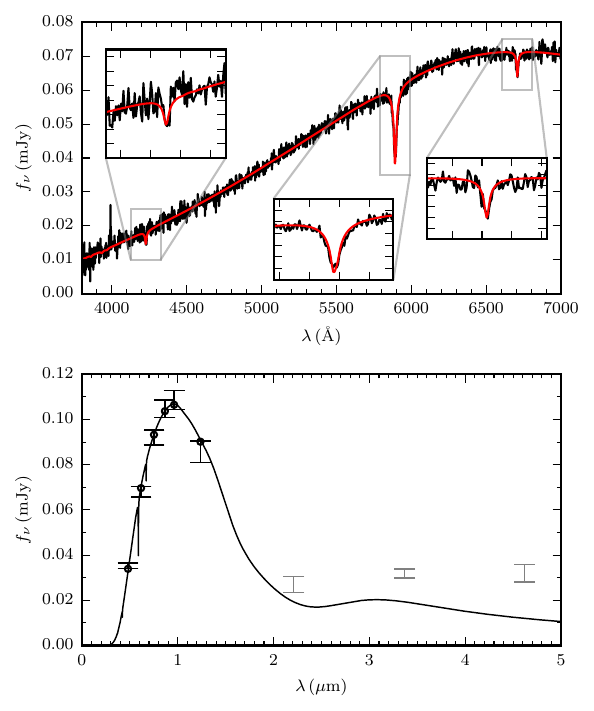}
\centering
\caption{WD~J2317+1830 model fit from Section \ref{sec:atm_modeling} to GTC/OSIRIS Data from \citet{Hollands2021_LiWD} and photometric points. The three longest wavelength photometric points are greyed out because they are ignored due to the infrared excess originating from a dust disk.  \label{fig:J2317_OSIRIS_model}}
\end{figure*}

\begin{figure*}[htb!]
\includegraphics[scale=1]{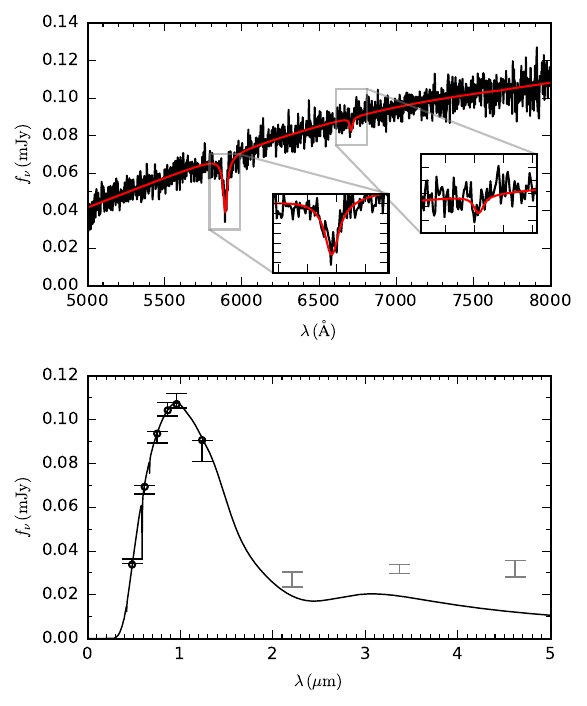}
\centering
\caption{WD~J2317+1830 model fit from Section \ref{sec:atm_modeling} to SOAR/Goodman Spectrograph Data from Section \ref{sec:obs} and photometric points. The three longest wavelength photometric points are greyed out because they are ignored due to the infrared excess originating from a dust disk.   \label{fig:J2317_goodman_fit}}
\end{figure*}

% \url{https://ctan.org/pkg/cjk?lang=en}.
%Further details on how to implement this and solutions for common problems,
%please go to \url{https://journals.aas.org/nonroman/}.


\bibliography{sample631}{}
\bibliographystyle{aasjournal}



\end{document}